**A Comprehensive Computational Photovoltaic Study of Lead-free Inorganic $NaSnCl_3$ -based Perovskite Solar Cell: Effect of Charge Transport Layers and Material Parameters**

*Md Tashfiq Bin Kashem*[a,1,*], *and Sadia Anjum Esha*[a,1]

[a] Electrical and Electronic Engineering, Ahsanullah University of Science and Technology, Dhaka – 1208, Bangladesh

[1] Equal contribution

[*] Corresponding author: E-mail address: tashfiq.eee@aust.edu

Lead-free all-inorganic halide perovskite solar cells (PSCs) have emerged as a promising alternative to toxic lead-based solar cells and organic solar cells, which have limited stability. This work explores such a PSC with sodium tin chloride ($NaSnCl_3$) as the absorber, due to its significant potential for optoelectronic applications. To investigate this potential, a comprehensive computational analysis of $NaSnCl_3$-based solar cells is performed using the one-dimensional solar cell capacitance simulator (SCAPS -1D). Simulations are performed for device structures with front contact/Indium Tin Oxide (ITO)/electron transport layer (ETL)/$NaSnCl_3$/hole transport layer (HTL)/back contact configuration, where $TiO_2$, $SnS_2$, IGZO, ZnSe, CdS, GaSe, $ZnSnN_2$, $WS_2$, PCBM, STO, and CSTO are utilized as ETLs and CNTS, GO, Mg-$CuCrO_2$, Spiro-OMeTAD, CdTe, GaAs, $MoTe_2$, $BaSi_2$, and P3HT are utilized as HTLs. Based on the obtained power conversion efficiency (PCE), six best ETL-HTL combinations with $SnS_2$, STO, $WS_2$, IGZO, ZnSe and CSTO as ETLs and $MoTe_2$ as HTL are chosen for further analysis. The effects of different material and device parameters, such as thickness and doping density; effective density of states; bulk and interface defects; series and shunt resistance; and operating conditions, such as temperature and light intensity are investigated. Using the optimized material parameters, $SnS_2$ ETL and $MoTe_2$ HTL-based solar cell show the best performance with open circuit voltage, $V_{oc}$ = 1.196V, short circuit current density, $J_{sc}$ = 35.82 mA/cm$^2$, fill factor, FF = 89.72% and PCE = 38.42%. This detailed study provides valuable insights for the fabrication of high efficiency $NaSnCl_3$-based solar cells.

## 1. Introduction

The industrial age has seen an increase in energy consumption faster than ever before due to population growth and technological development. The ever-growing demand for energy is hard to meet from non-renewable sources (coal, natural gas, and oil) alone, mainly due to their depletion rates and negative impacts on the environment, such as rapid climate change, air pollution, global warming [1]. The challenge can be addressed by using renewable energy sources namely solar energy, wind energy, hydropower, biomass energy, tidal energy etc. [2] Solar photovoltaic (PV) cell that utilize solar energy, the most abundant renewable energy source, is considered one of the most effective solutions due to its ease of maintenance, scalability to meet growing demand, and ability to be installed in existing building structures around the world [3].

The first-generation solar cells include photovoltaic cells (PVCs) with thick mono- and poly-crystalline silicon-based films. Although these cells formed the basis for commercial solar devices, they could not be very useful because of having low efficiency and high manufacturing cost [4], [5].

With the advancement of time and technology, the second generation of solar cells were developed that were made using thin films of cadmium sulfide (CdS), cadmium telluride (CdTe), copper indium gallium selenide (CIGS) or amorphous silicon. These thin-film technologies provided comparatively better mechanical characteristics and thus could be integrated into flexible applications. But the efficiencies still remained low, even lower PCEs were spotted at times in thin film solar cells [6].

The transformation to third-generation photovoltaic technologies, especially perovskite solar cells, is one of the most significant advancements. Perovskite materials, used as absorber in PSCs, have the general chemical formula $ABX_3$, where A is monovalent cation (organic cation such as $CH_3NH_3{}^+$, $CH_3CH_2NH_3{}^+$, $NH_2CHNH_2{}^+$ [7], [8] or alkali cation such as Cs+, Rb+, K+, Na+, and Li+), B is divalent cation ($Pb^{2+}$, $Sn^{2+}$, $Ge^{2+}$) and X is an anion (oxygen, nitrogen, carbon or halogen) [8], [9]. They possess some unique optical and electrical properties including tunable bandgap, large optical absorption coefficient, high carrier diffusion length and lifetime [8], [10], [11]. Combining remarkable PCEs, and low production costs, perovskite solar cells have become one of the most promising solar technologies. However, one of the best performing PSCs utilize lead (Pb) for the B element in $ABX_3$ perovskite, which involve environmental pollution and health risks because of the toxicity of lead, posing an obstacle for mass commercialization [12]–[14]. On the other hand, the presence of organic cations in perovskite such as methylammonium ($CH_3NH_3^+$), formamidinium ($CH(NH_2)_2^+$) makes the material thermodynamically unstable as they cannot withstand high temperature, resulting in shorter lifetime of the corresponding devices [15], [16].

A number of studies have been conducted utilizing non-toxic elements to replace lead and inorganic cations to substitute organic ones, to get rid of the toxicity issue and improve the stability of PSCs [17]–[20]. As non-toxic and environment-friendly metals, tin (Sn), germanium (Ge), bismuth (Bi), antimony (Sb), and copper (Cu) have been explored [21]–[30]. Among all these lead-free alternatives, Sn-based perovskites have shown the greatest potential due to their attractive material properties similar to those of Pb-based perovskites [13], [31]. Being the neighbor of Pb in group 14 of the periodic table, Sn has similar outer electronic configuration as Pb ($ns^2$ $np^2$). Smaller ionic radius of $Sn^{2+}$ (1.35 Å) compared to $Pb^{2+}$ (1.49 Å), enables Sn to replace Pb in the perovskite structure without causing significant distortion [31]. Moreover, compared to Pb-based counterparts, Sn-based perovskites show superior properties such as narrower bandgap and higher charge carrier mobility [31]–[33].

On the other hand, different alkali metals (Cs, Rb, K and Na) have been utilized in place of organic cations for the A site of perovskites and better thermal stability with high efficiency have been reported [34]–[38]. Therefore, lead-free all-inorganic perovskites have found to be an efficient alternative to organic-inorganic lead-based perovskites [39]–[41]. In this work, one such perovskite, with high potential for PSC, namely sodium-tin-chloride ($NaSnCl_3$) is investigated through computational modeling of solar cells. $NaSnCl_3$ has been reported to have exciting properties suitable for optoelectronic technologies such as narrow bandgap, high absorption coefficient, low reflectivity in the visible to infrared range, suitable refractive index and high optical conductivity across a wide photon energy range [42]–[46]. However, limited studies have been reported till now on the photovoltaic performance of $NaSnCl_3$ in solar cells [47], to the best of the authors' knowledge.

Bouri *et al.* investigated $NaSnCl_3$ and $KSnCl_3$ absorber based perovskite solar cells, with $TiO_2$ as the ETL and Spiro-OMeTAD as the HTL [47]. They investigated the effects of various parameters of the absorber such as thickness, donor and acceptor concentrations, and conduction and valence band effective density of states, on the device performance and reported a PCE of 36.14% and 34.56% for $NaSnCl_3$ and $KSnCl_3$ respectively. However, further simulations involving other possible ETLs and HTLs with better band alignment with the absorber could provide valuable insights and potentially higher PCE with optimized material parameters. Additionally, Spiro-OMeTAD has certain disadvantages as a HTL. It is costly, prone to instability because of its degradable organic components [48], [49], and can accelerate the breakdown of the PSC, reducing the life time of the cell [50].

In this work, a broad analysis of $NaSnCl_3$ absorber based perovskite solar cell is performed using Al (Aluminum, front contact)/ITO (Indium-Tin-Oxide, window layer)/ETL/$NaSnCl_3$ (absorber)/HTL/Au (Gold, back contact) structure. First, simulations are performed with ninety-nine (99) possible combinations from eleven (11) ETLs and nine (9) HTLs to determine the most efficient device configurations. The ETLs considered include $TiO_2$, $SnS_2$, IGZO, ZnSe, CdS, GaSe, $ZnSnS_2$, $WS_2$, PCBM, STO and CSTO. The HTLs tested include P3HT, $BaSi_2$, $MoTe_2$, GaAs, CdTe, Spiro-OMeTAD, Mg–$CuCrO_2$, GO, and CNTS. Six combinations with ETLs ─ $SnS_2$, STO, $WS_2$, IGZO, ZnSe, and CSTO and HTL ─ $MoTe_2$ show higher PCE and these are further utilized to study the effects of the thickness and doping density of the absorber, as well as those of the ETL and HTL. Furthermore, the effects of series resistance, shunt resistance, temperature, light intensity, conduction and valence band effective density of states in the absorber, and the defect density in the absorber and at the absorber/ETL and absorber/HTL interfaces are explored. Finally, simulations are performed with the optimized material parameters to determine the maximum possible efficiency of the devices.

## 2. Simulation methodology

SCPAS-1D software (version 3.3.11) is utilized for the simulations of this work. This numerical simulation tool was developed by Burgelman *et al.* from the Electronics and Information Systems (ELIS) department of University of Gent to model the photovoltaic characteristics of thin film solar cells [51]. Results obtained using this software demonstrate excellent match with the experimental outcomes, as reported in previous studies [52]–[59]. SCAPS-1D simultaneously solves Poisson equation, continuity equations for charge carriers, and drift-diffusion based charge transport equations, as described by equations (1) – (5), to calculate the PV performance parameters [17], [60].

Poisson equation:

$$\frac{\partial^2 \Phi(x)}{\partial^2 x} = -\frac{q}{\epsilon}\left[p(x) - n(x) + N_D^+ - N_A^- \pm N_{def}(x)\right] \quad (1)$$

Continuity equation:

$$\begin{cases} \text{Electron: } \dfrac{\partial n}{\partial t} = \dfrac{1}{q}\dfrac{\partial J_n}{\partial x} + (G_n - R_n) & (2) \\ \text{Hole: } \dfrac{\partial p}{\partial t} = -\dfrac{1}{q}\dfrac{\partial J_p}{\partial x} + (G_p - R_p) & (3) \end{cases}$$

Charge transport equation:

$$\begin{cases} \text{Electron: } J_n = qn\mu_n E + qD_n \dfrac{\partial n}{\partial x} & (4) \\ \text{Hole: } J_p = qp\mu_p E - qD_p \dfrac{\partial p}{\partial x} & (5) \end{cases}$$

Here, $\Phi$ is the electrostatic potential, $q$ is the electronic charge, $\epsilon$ is the permittivity of the material, $p$ and $n$ are hole and electron density respectively, $N_D^+$ and $N_A^-$ are donor and acceptor density respectively, $N_{def}$ is the defect density, $G$ and $R$ are generation and recombination rate respectively, $J$ is the current density, $E$ is the electric field and D is the diffusion coefficient. In all the equations, the subscripts, $n$ and $p$ correspond to electron and hole respectively. Additionally, to compute the optical absorption in each layer, the following 'Eg-sqrt' model is utilized in SCAPS-1D [61]:

$$\alpha(h\upsilon) = \begin{cases} \left[\alpha_0 + \beta_0 \frac{E_g}{h\upsilon}\right] \sqrt{\frac{h\upsilon}{E_g} - 1}, & E_g < h\upsilon \\ 0, & E_g > h\upsilon \end{cases} \tag{6}$$

Here, α is the absorption coefficient of the material, $\alpha_0$ and $\beta_0$ are model constants, $E_g$ is the energy bandgap of the material, hυ is photon energy where h is Planck's constant, and υ is photon frequency.

PCE and FF are calculated using the following equations,

$$PCE = \frac{V_{oc} J_{sc} FF}{P_{in}} \tag{7}$$

$$FF = \frac{V_m J_m}{V_{oc} J_{sc}} \tag{8}$$

Where $V_m$ and $J_m$ correspond to the voltage and current density respectively at the maximum power point, and $P_{in}$ is the input power.

### 3. Device structure and material parameters

The solar cell structure studied for this work consists of the following configuration: front contact/ITO/ETL/$NaSnCl_3$/HTL/back contact, as schematized in Fig. 1a. Aluminum having work function of 4.2 eV [62] is used as the front contact metal, with ITO as the transparent conducting oxide (TCO). $NaSnCl_3$ works as the absorber or light harvester material, where electron-hole pairs are generated upon the absorption of light (Fig. 1b). ETL and HTL act as the charge transport layers. ETL collects the generated electrons, and transports them to the front contact metal (cathode). Besides, ETL blocks the hole transport from the absorber layer. On the other hand, HTL facilitates the transport and collection of generated holes to the back contact metal (anode) while blocking the electron transport from the absorber layer. Gold with work function of 5.1 eV [63] is chosen as the back contact metal for this work. Table 1 and Table 2 show the initial material parameters of different layers. Unless otherwise stated, all simulations are performed under AM1.5G sunlight spectrum with incident light power of 1000 $W/m^2$ and temperature of 300K.

## 4. Results and discussion

### 4.1 Selection of optimal ETL-HTL combinations for high efficiency

Initially, simulations are performed using all 99 possible combinations out of 11 ETLs and 9 HTLs, as listed in Table 1 and 2, to compare and identify the most efficient device configurations. The corresponding PV performance parameters are shown in Fig. 2─5. Higher $V_{oc}$ (≥ 0.47 V) is obtained with $BaSi_2$ as HTL, using $SnS_2$, STO, GaSe, $WS_2$, IGZO, PCBM, CdS, ZnSe and CSTO as ETLs. Higher $J_{sc}$ (≥ 42.8 $mA/cm^2$) is achieved with $MoTe_2$ as the HTL and any of the 11 ETLs, except $ZnSnN_2$. Poor FF (below 50%) is observed with $BaSi_2$, and GO as HTLs, regardless of the ETL choice. The highest FF (~ 70%) is achieved with CSTO or ZnSe as ETL and $MoTe_2$ as HTL. Poor PCE (~ 4 %) is obtained with $BaSi_2$ HTL, regardless of the ETL selection. Devices with $MoTe_2$ as HTL and any of these six ETLs ─ $SnS_2$, STO, $WS_2$, IGZO, ZnSe and CSTO ─ show higher PCEs of ~14%. Therefore, these six structures are identified as the best-performing devices (performance parameters are listed in Table 3) and utilized in subsequent studies to investigate the effects of varying material parameters on device performance. The energy band diagrams of the devices are shown in Fig. 6.

### 4.2 Effect of acceptor and donor density of the absorber

The effect of the donor and acceptor concentrations of the absorber material on the PCE is studied for six structures (Fig. 7). In most cases, it is observed that the PCE is higher for a higher acceptor concentration and a lower donor concentration. After being photo-generated, carriers move toward the depletion regions of the heterojunctions (absorber/ETL and absorber/HTL), where they are separated by the built-in electric field. As the acceptor concentration increases, the width of the depletion region decreases, and the electric field increases. A narrower depletion region leads to a lower number of carriers being separated and collected due to the higher probability of carrier recombination, whereas the increased electric field enhances carrier collection at the respective electrodes [64]. The findings of this study suggest that the latter effect plays a more dominant role.

## 4.3 Effective density of states in the conduction band and valence band of the absorber

Effective density of the states in the conduction band ($N_c$) and valence band ($N_v$) of the absorber material significantly influence device performance, as evidenced by the plots shown in Fig. 8─11. Table 4 summarizes the range of $N_c$ and $N_v$ required to achieve higher $V_{oc}$, $J_{sc}$, FF and PCE across all six structures. Higher $V_{oc}$ can be achieved when both $N_c$ and $N_v$ are lower (within the range of $10^{16}$ $cm^{-3}$ to $10^{17}$ $cm^{-3}$), higher $J_{sc}$ is obtained for higher $N_c$ in the range $10^{18}$ $cm^{-3}$ to $10^{19}$-$10^{20}$ $cm^{-3}$ and higher $N_v$ in the range $10^{19}$ $cm^{-3}$ to $10^{20}$ $cm^{-3}$, larger FF can be achieved with lower $N_c$ in the range $10^{16}$ $cm^{-3}$ to $10^{17}$ $cm^{-3}$ (except $WS_2$ and CSTO ETL-based structures, where higher $N_c$ is required) and lower $N_v$ in the range $10^{16}$ $cm^{-3}$ to $10^{17}$ $cm^{-3}$, and higher PCE corresponds to lower $N_c$ ($10^{16}$ $cm^{-3}$) and lower $N_v$ ($10^{16}$ $cm^{-3}$). The findings are consistent with the analytical and numerical studies reported by Zhou *et al.* for various perovskite solar cells [65]. Lower $N_c$ and $N_v$ correspond to a smaller effective mass, which enhances carrier mobility and contributes to higher efficiency in the solar cell.

## 4.4 Effect of absorber thickness

The effects of varying absorber thickness on $V_{oc}$, $J_{sc}$, FF and PCE for six structures are presented in Fig. 12 and 13. $V_{oc}$ declines with increasing absorber thickness, which can be attributed to the reduced electric field that results in a decrease in the transport of photo-generated carriers to the corresponding electrodes [66]. Besides, the increased dark saturation current and carrier recombination contribute to the continuous decrease in $V_{oc}$ up to a certain absorber thickness, above which it becomes constant [67]. From another perspective, a thicker absorber layer results in more photon absorption, leading to an increase in electron-hole pair generation. However, it also creates a longer path for the photo-generated charge carriers to travel to the electrodes, increasing recombination [68]. In this study, the latter effect is found to become dominant, as $J_{sc}$ decreases with increasing absorber thickness in all six structures. Increased recombination also results in the decrease in FF with increasing absorber thickness [64]. However, in the structure with $WS_2$ as ETL, the decrease in sheet resistance with the increase in absorber thickness becomes more prominent, leading to an increase in FF [69]. Again, due to the reduced electric field and the increased recombination within the absorber layer, carrier transport is hindered, leading to a decrease in PCE as the absorber thickness increases [69].

### 4.5 Effect of ETL thickness and doping density

The effects of the thickness (0.1 μm to 0.5 μm) and doping density ($10^{15}$ $cm^{-3}$ to $10^{20}$ $cm^{-3}$) of ETL on device performance are investigated (Fig. 14–17). In $WS_2$, IGZO, ZnSe and CSTO ETL-based structures, higher $V_{oc}$ is achieved with a thicker ETL (0.4 μm – 0.5 μm) and lower doping density ($10^{15}$ $cm^{-3}$). In $SnS_2$ and STO ETL-associated structures, $V_{oc}$ is high and almost independent of ETL thickness and doping density for a large range, it starts decreasing beyond a doping density of $10^{19}$ $cm^{-3}$. Higher $J_{sc}$ is observed for a higher doping density of ETL. The $SnS_2$ ETL-based structure shows only minor variation in $J_{sc}$ with changes in ETL thickness or doping density; for structures with STO, $WS_2$ and ZnSe as ETLs, higher $J_{sc}$ is independent of ETL thickness when a high doping density ($\geq 10^{19}$ $cm^{-3}$) is used. In the CSTO as ETL-associated structure, higher $J_{sc}$ is independent of doping density as long as the ETL thickness remains between 0.1 μm and 0.3 μm. FF in the $SnS_2$ as ETL-based structure remains independent of ETL thickness and doping density over a wide range (0.1 μm to 0.5 μm and $10^{15}$ $cm^{-3}$ to $10^{19}$ $cm^{-3}$), and decreases slightly when the doping density exceeds $10^{19}$ $cm^{-3}$. For the IGZO, ZnSe, and CSTO as ETL-based structures, higher FF is achieved when a higher doping density ($\geq 10^{19}$ $cm^{-3}$) is used. In the STO as ETL-associated structure, higher FF is obtained with ETL thickness in the range of 0.3 μm to 0.5 μm and lower doping density of ETL ($\sim 10^{16}$ $cm^{-3}$). Higher FF can be obtained in the $WS_2$ ETL-based structure at any ETL thickness within the studied range when the ETL doping density is between $10^{15}$ $cm^{-3}$ and $10^{17}$ $cm^{-3}$. Higher PCE is independent of ETL thickness when a higher doping density is used ($10^{18}$ $cm^{-3}$ – $10^{19}$ or $10^{20}$ $cm^{-3}$ for IGZO and CSTO as ETLs and above $10^{19}$ $cm^{-3}$ for STO, $WS_2$ and ZnSe as ETLs). In the $SnS_2$ ETL-based structure, PCE variation is minimal, with a slightly higher value is observed for low-to-moderate doping density ($10^{15}$ $cm^{-3}$ to $10^{18}$ $cm^{-3}$) and thicker ETL (0.4 μm to 0.5 μm). A much thicker ETL layer may lead to increased parasitic absorption, reduced transmittance and fewer electron-hole pairs generated in the absorber layer, resulting in inferior performance parameters [70][71]. On the other hand, higher doping density in the ETL material leads to an increase in the built-in electric field at the ETL/absorber junction, which enhances the drift of majority carriers while creating a larger barrier for minority carriers [50], [72]. As a result, recombination is expected to decrease, leading to an increase in device efficiency.

### 4.6 Effect of HTL thickness and doping density

Device performance is analyzed over a range of HTL thickness (0.1 μm to 0.5 μm) and doping density ($10^{15}$ $cm^{-3}$ to $10^{20}$ $cm^{-3}$) as shown in Fig. 18–21. Thicker HTL (0.4 μm – 0.5 μm) and higher doping density ($10^{18}$ $cm^{-3}$ – $10^{20}$ $cm^{-3}$) lead to higher performance parameters in all structures, though the variations are small. An appropriately thick HTL can improve light absorption in the device, thereby enhancing quantum efficiency [73]. It also facilitates efficient transport of photo-generated holes to the back contact with reduced recombination, leading to greater charge collection efficiency [74]. However, an excessively thick HTL can lead to increased recombination and higher resistance, thereby reducing solar cell performance [66], [74]. A higher acceptor concentration in the HTL increases its conductivity [72]. Furthermore, it leads to a stronger built-in electric field at the HTL/ absorber junction, which enhances the drift of photo-generated holes and repels the associated electrons. This reduces recombination at the interface and improves efficiency [75].

### 4.7 Effect of defects in the absorber

Defects in the absorber layer, including point defects, surface defects, and dopant-related defects, can significantly impact the performance of solar cells by acting as trap sites and potential recombination centers. Consequently, the trap-assisted non-radiative recombination process known as Shockley-Read-Hall (SRH) recombination, increases, resulting in a reduction in carrier mobility, lifetime and diffusion length [74], [76], [77], and an ultimate decrease in efficiency. In this work, the defect density of $NaSnCl_3$ is varied from $10^{12}$ $cm^{-3}$ to $10^{18}$ $cm^{-3}$ (Fig. 22). Both $V_{oc}$ and PCE consistently decrease as defect density increases. $J_{sc}$ remains constant initially but begins to decrease when the defect density exceeds $10^{14}$ $cm^{-3}$. However, in the $WS_2$ as ETL-based structure, $V_{oc}$ and PCE become saturated beyond the defect density of $10^{15}$ $cm^{-3}$ and $J_{sc}$ initially decreases with increasing defect density but becomes saturated when the defect density exceeds $10^{15}$ $cm^{-3}$. FF decreases slowly up to a defect density of $10^{14}$ $cm^{-3}$, after which the rate of decrease becomes more pronounced. However, in the STO as ETL-based structure, FF becomes saturated if the defect density becomes higher than $10^{16}$ $cm^{-3}$ and in the device with $WS_2$ as ETL, FF initially decreases up to a defect density of $10^{14}$ $cm^{-3}$, then increases and becomes saturated when the defect density exceeds $10^{16}$ $cm^{-3}$.

### 4.8 Defects at the ETL/absorber and HTL/absorber interfaces

The effects of interface defect density, varying in the range from $10^{10}$ $cm^{-2}$ to $10^{18}$ $cm^{-2}$, at the ETL/absorber and HTL/absorber interfaces, are demonstrated in Fig. 23─26. Typically, the effect of interface defects on the photovoltaic parameters is minimal up to a certain defect density (around $10^{12}$ $cm^{-2}$ – $10^{14}$ $cm^{-2}$), beyond which the parameters begin to decrease. A high defect density leads to the formation of more traps and recombination centers, resulting in increased SRH recombination, reduced carrier lifetime, and a decline in performance parameters [75], [78]. As evident from the figures, the HTL/absorber interface defect has a smaller effect on device performance compared to the ETL/absorber interface. Furthermore, all performance parameters are found to saturate above an HTL/absorber defect density of approximately $10^{16}$ $cm^{-2}$. FF in the structure with $WS_2$ as the ETL shows an increasing trend with higher defect density for both ETL/absorber and HTL/absorber interfaces.

### 4.9 Series and shunt resistance

Series resistance ($R_s$) and shunt resistance ($R_{sh}$) characterize loss mechanisms within the solar cell and can have a substantial effect on the cell performance if they exceed a certain threshold. $R_s$ and $R_{sh}$ can be represented in the equivalent circuit of a solar cell, as shown in Fig. 27, where I is the current delivered to the load, $I_L$ is the photocurrent, $I_D$ is the dark current and $I_{sh}$ is the current through the shunt resistance [79]. $R_s$ is primarily determined by the interfaces between ETL/absorber and HTL/absorber, the contact between the charge transport layers (ETL and HTL) and the metal contact layers, as well as the resistance of the front and back metal contacts. These factors collectively influence charge transfer [63], [73], [80], [81]. A higher series resistance indicates an increased barrier to the transport of photo-generated charge carriers, and leads to greater power loss [82][83]. In this work, $R_s$ is varied from 0 to 6 $\Omega.cm^2$ across all six structures, as shown in Fig. 28. It is evident that $V_{oc}$ and $J_{sc}$ remain unchanged in the entire range. FF decreases significantly with an increase in $R_s$, indicating less output power at higher resistance. As a result, PCE decreases, with the reduction being more pronounced in structures with $SnS_2$, IGZO, ZnSe and CSTO as the ETL.

On the other hand, the primary source of shunt resistance is manufacturing defects. Low shunt resistance results in power losses in solar cells by creating an alternate pathway for the light-generated current [81]. In this study, $R_{sh}$ is varied from 10 $\Omega.cm^2$ to $10^7$ $\Omega.cm^2$. $J_{sc}$ remains nearly constant across this range, while $V_{oc}$ drops if $R_{sh}$ falls below $10^2$ $\Omega.cm^2$. FF and PCE decline when $R_{sh}$ drops below $10^3$ $\Omega.cm^2$ due to increased power loss. Several techniques have been reported to achieve high shunt resistance such as, using solvent additives to improve morphological properties by preventing pinholes or voids and adding interface layer made of passivation materials to reduce the shunt current [63], [84]–[87].

### 4.10 Effect of temperature

The effects of temperature on the performance parameters are shown in Fig. 29 and 30. $V_{oc}$, FF and PCE decrease, while $J_{sc}$ increases with increasing temperature. As temperature rises, the bandgap of most semiconductors narrows [75], [88], [89], leading to a higher generation of photo-induced carriers and, consequently, an increase in $J_{sc}$. However, the smaller bandgap at elevated temperatures also enhances the thermal generation of electron-hole pairs, resulting in higher dark currents and a reduction in $V_{oc}$. Besides, the recombination of charge carriers accelerates due to the reduction in bandgap, thermal excitation of carriers and activation of various trap sites as temperature increases [75], ultimately leading to a reduction in FF and PCE. However, temperature rise results in increased FF and PCE for some ETLs at few temperatures (Fig. 30b,c,e,f), which can be attributed to the dominance of the increased photo-generation of electron-hole pairs.

### 4.11 Effect of incident light intensity

The six structures are analyzed under different light intensities to investigate their potential for both outdoor use and indoor applications (lower incident irradiance) and the results are shown in Fig. 31. $J_{sc}$ increases linearly with increasing light intensity, and this increase is quite significant, which can be attributed to the higher input power and the increased generation of photo-carriers [90]. $V_{oc}$ increases logarithmically with light intensity, showing a steep rise up to an intensity of 400 W/m$^2$, followed by a gradual increase at higher intensities. FF increases slightly with light intensity due to the counteracting interplay between the increased photo-generation of carriers and the enhanced carrier recombination [90]. The latter effect appears to dominate in the structure with CSTO as the ETL, as evidenced by a slight decrease in FF with increasing light intensity. As a result of the combined effects of a significant increase in $J_{sc}$, a moderate increase in $V_{oc}$, a slight increase in FF, and the increased input power, PCE demonstrates a slow increase with light intensity for all of the structures.

### 4.12 Optimized performance of the devices

Finally, PV performance metrics of the six structures are optimized for 300K temperature and 1000 W/m$^2$ incident light intensity, utilizing the following parameters of the absorber: $N_c = N_v = 10^{16}$ cm$^{-3}$, $N_a = 10^{20}$ cm$^{-3}$, $N_d = 10^{15}$ cm$^{-3}$, and thickness = 1 μm while other key parameters are set as follows: ETL and HTL thickness = 0.4 μm, defect density in the absorber, ETL and HTL = $10^{15}$ cm$^{-3}$, ETL/absorber and HTL/absorber interface defect density = $10^{12}$ cm$^{-2}$. Current-voltage characteristics of the optimized device structures are shown in Fig. 32a. $V_{oc}$ are almost the same for all structures. Relatively lower $J_{sc}$ is observed for $WS_2$ and STO ETL-based structures. All the values are listed in Table 5.

Corresponding quantum efficiency (QE) [81] of the optimized structures as a function of the wavelength of incident photons are presented in Fig. 32b. For all the structures, QE decreases when the wavelength is below ~360 nm, which can be attributed to light absorption and recombination at the front surface [81]. STO ETL-based structure shows low QE, $WS_2$ ETL-based structure shows higher QE up to ~690 nm, after which it decreases, even falling below that of the STO as ETL case. IGZO, ZnSe and CSTO ETL-based structures show almost similar QE behavior; compared to these, $SnS_2$ ETL-based structure demonstrates higher QE up to wavelength of ~560 nm, beyond which, it shows similar QE to the others.

Table 5 presents the PV performance parameters for the six optimized structures in this work. Compared to the device structure with the same absorber material reported in ref. [47], higher $V_{oc}$ is obtained in all the six structures of this study and higher $J_{sc}$, FF and PCE are achieved in $SnS_2$, IGZO and ZnSe ETL-based structures. The best ETL-HTL combination based on this work is found to be $SnS_2$-$MoTe_2$, with $V_{oc}$ = 1.196 V, $J_{sc}$ = 35.82 mA/cm$^2$, FF = 89.72% and PCE = 38.42%. Slightly higher $V_{oc}$ of 1.202 V can be achieved with $WS_2$ as ETL, though the other parameters are found to be comparatively smaller.

## 5. Conclusion

This work focuses on the detailed investigation of lesser-studied, yet highly promising halide perovskite, $NaSnCl_3$-based solar cell. 11 materials as ETL and 9 materials as HTL are explored. Depending on the electron affinity, energy bandgap and doping density of the materials, different types of band alignments occur at the interfaces between absorber and ETL/HTL, which affect charge transfer and PV characteristics. Six structures with ETLs – $SnS_2$, STO, $WS_2$, IGZO, ZnSe and CSTO, and HTL – $MoTe_2$ result in higher PCE and these devices are further explored to study the impact of different device and material parameters on device performance and to optimize the parameters for the best output. Using the optimized material parameters, the best performance is achieved for $SnS_2$ ETL and $MoTe_2$ HTL-based structure which demonstrated $V_{oc}$ of 1.196 V, $J_{sc}$ of 35.82 mA/cm$^2$, FF of 89.72% and PCE of 38.42%, which are higher than the previously reported result for $NaSnCl_3$ absorber based solar cell. The findings of this work highlight the significant potential of environment-friendly lead-free all-inorganic $NaSnCl_3$-based perovskite solar cell and pave the way for advancing sustainable solar energy technologies.

## Acknowledgements

The authors would like to convey their sincere gratitude to Dr. M. Burgelman of the University of Gent in Belgium for kindly providing the SCAPS-1D software.

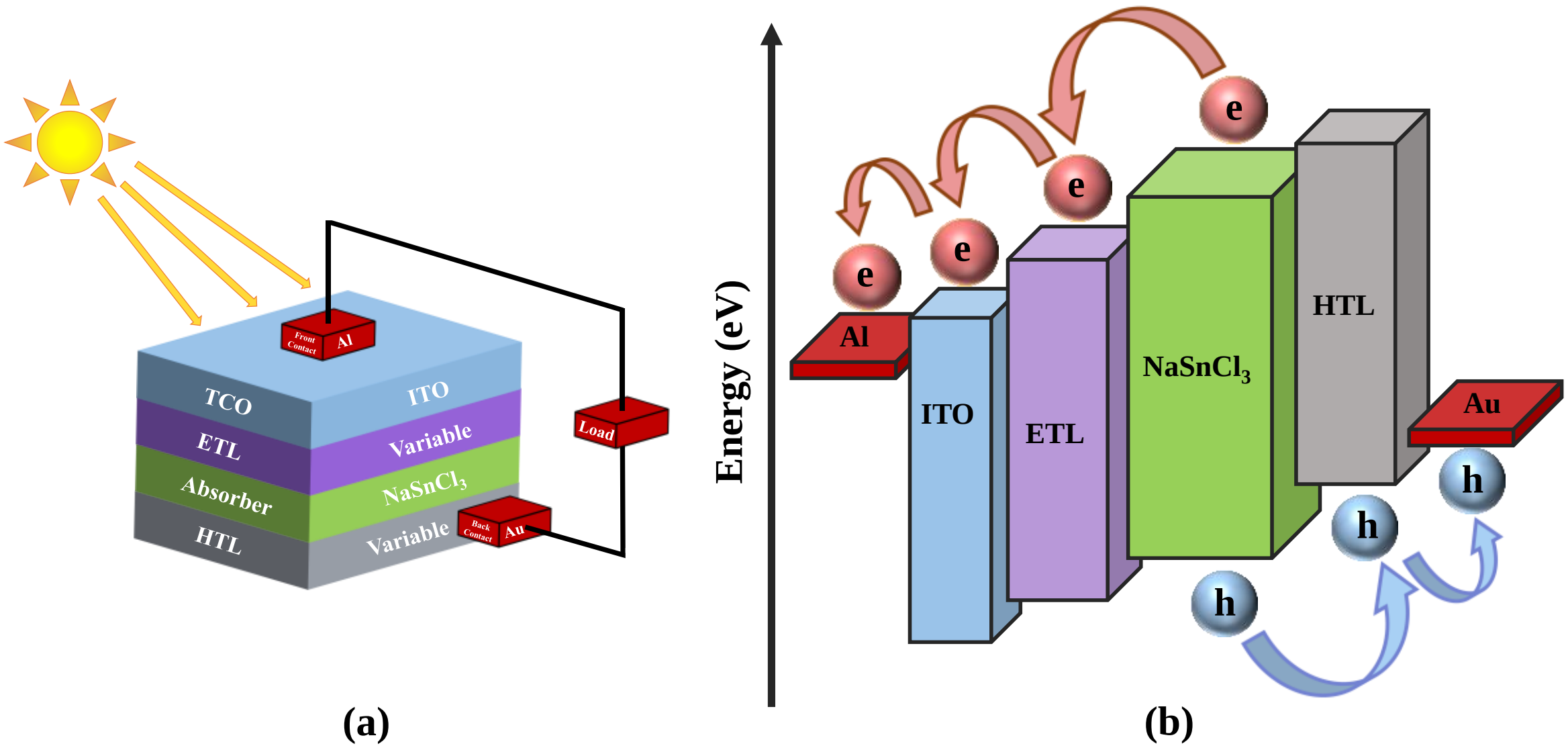

Fig. 1. Schematic diagram of the (a) perovskite solar cell configuration used in simulations, and (b) energy levels of different layers. Variable term for ETL and HTL in (a) refer to 11 different ETL and 9 different HTL materials explored in this study. ‘e’ and ‘h’ inside the sphere in (b) represent electron and hole respectively.

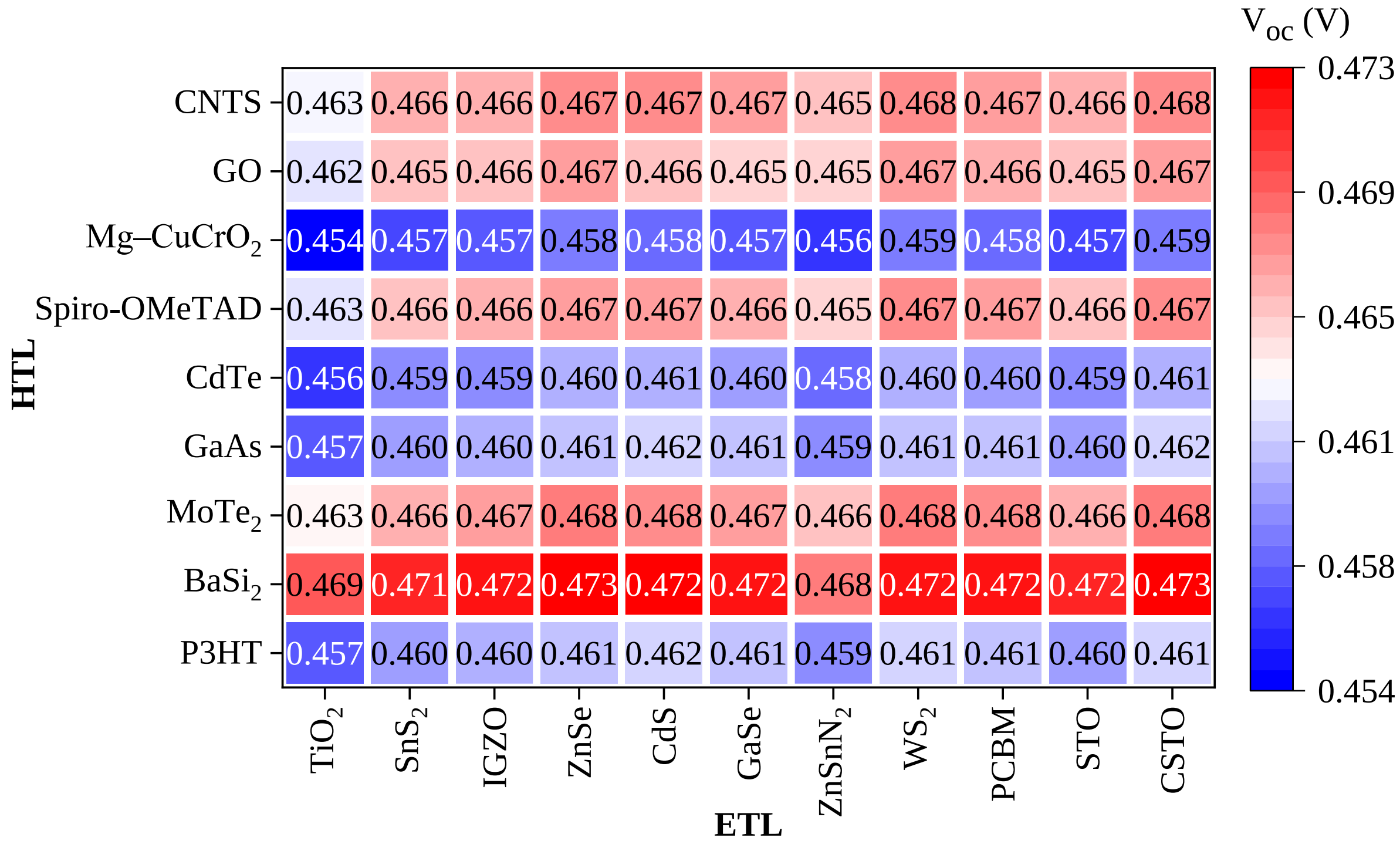

Fig. 2. $V_{oc}$ for $NaSnCl_3$-based solar cell structures with different ETLs and HTLs. Al and Au are used as front and back contact metal and ITO is used as TCO.

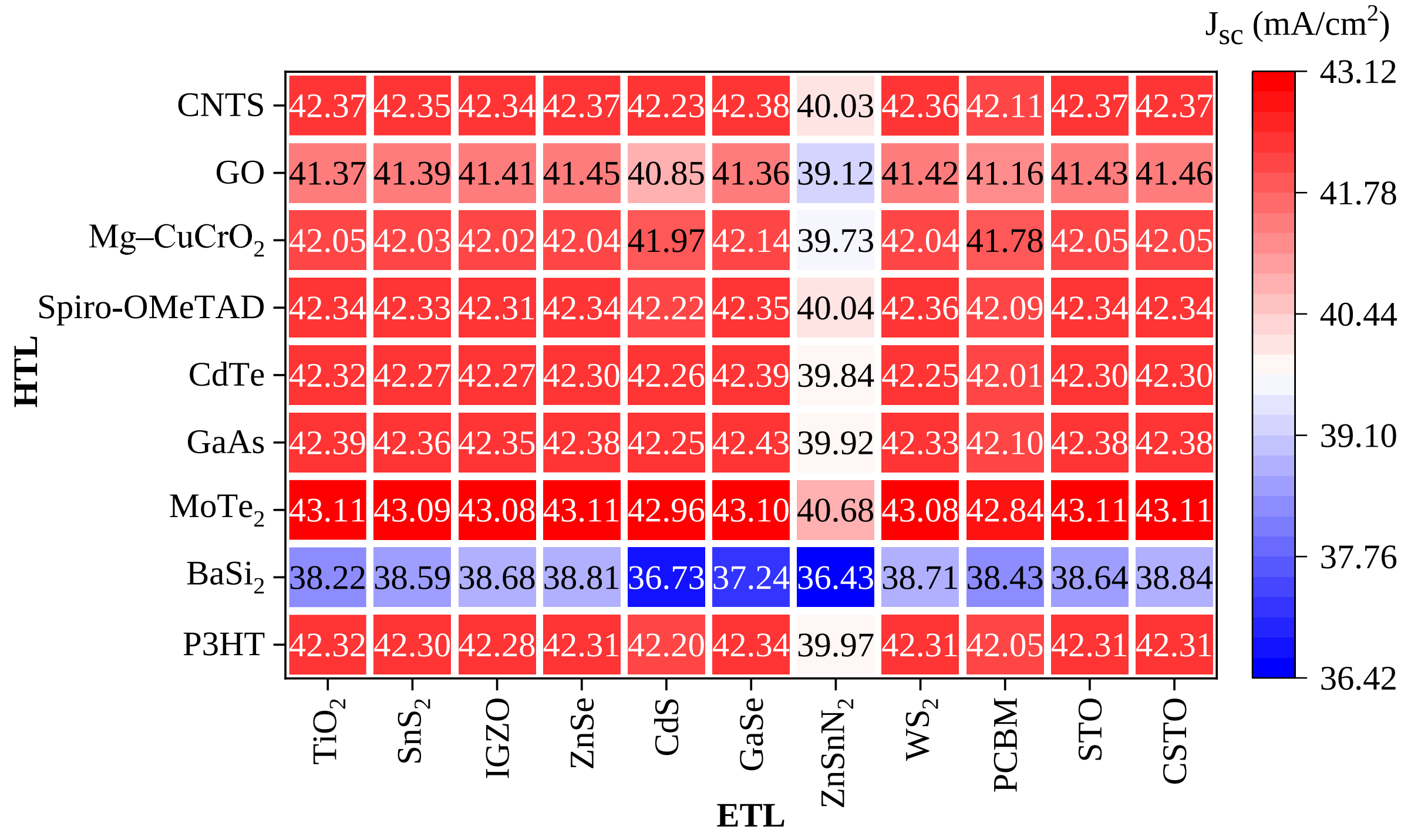

Fig. 3. $J_{sc}$ for $NaSnCl_3$-based solar cell structures with different ETLs and HTLs. Al and Au are used as front and back contact metal and ITO is used as TCO.

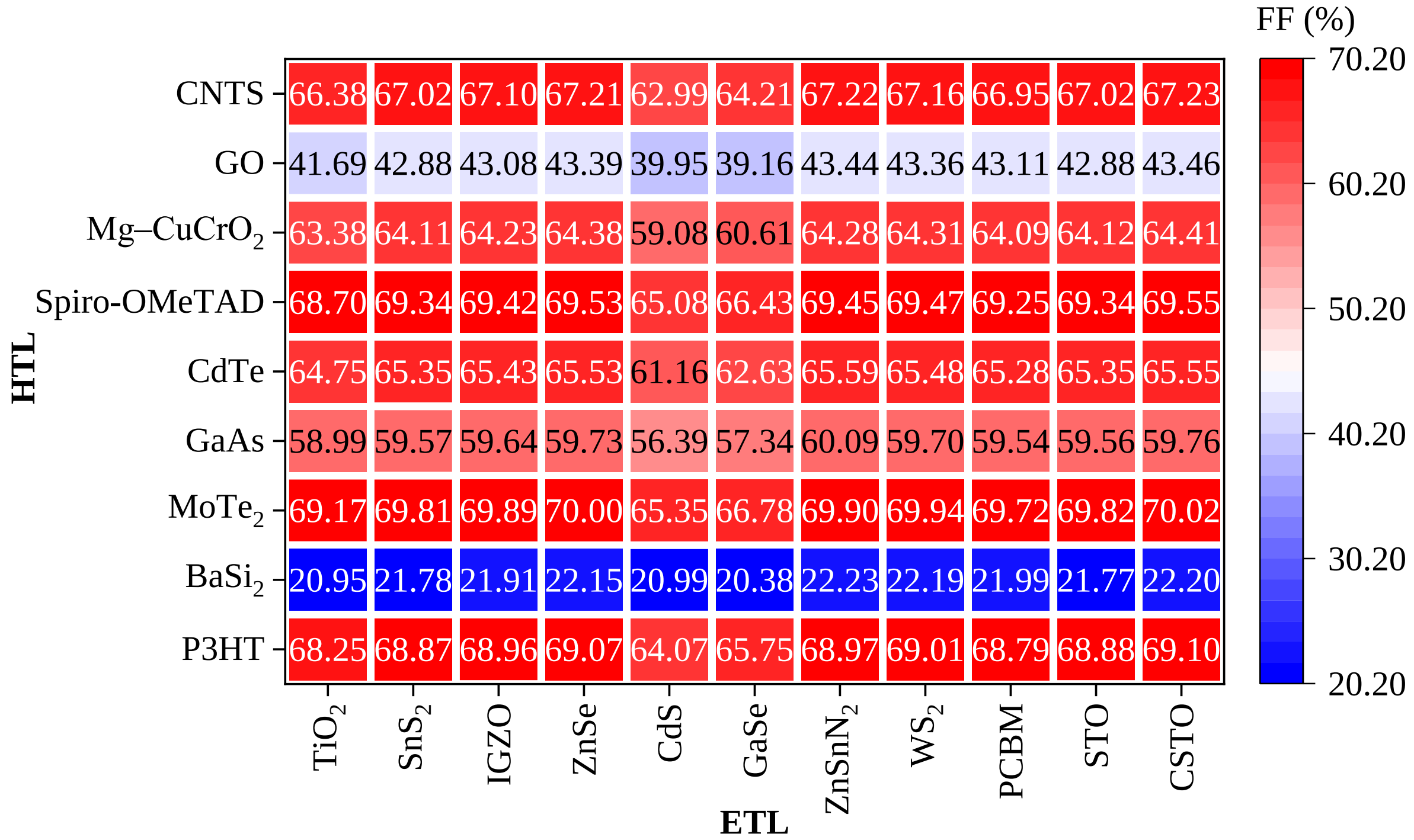

Fig. 4. FF for $NaSnCl_3$-based solar cell structures with different ETLs and HTLs. Al and Au are used as front and back contact metal and ITO is used as TCO.

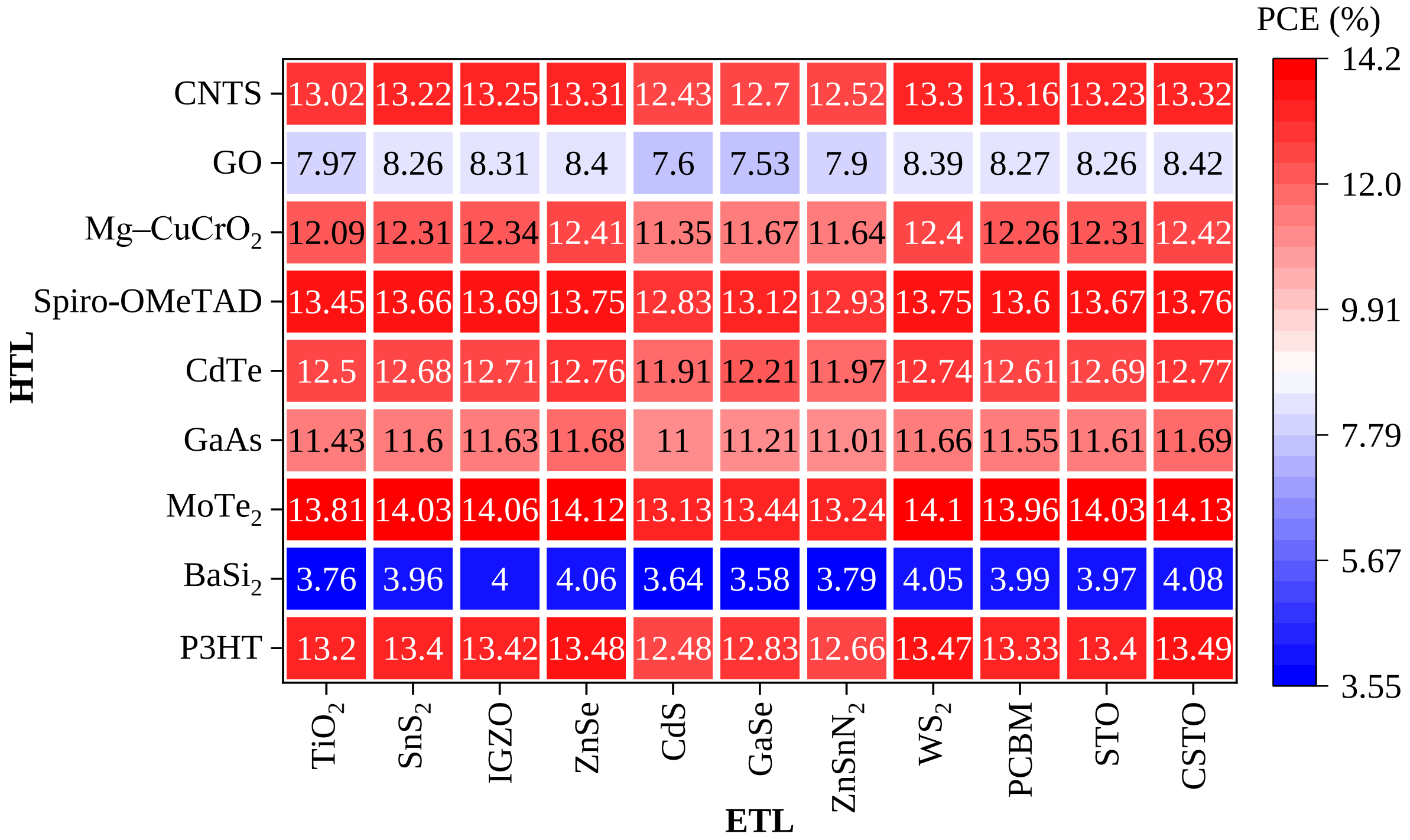

Fig. 5. PCE for $NaSnCl_3$-based solar cell structures with different ETLs and HTLs. Al and Au are used as front and back contact metal and ITO is used as TCO.

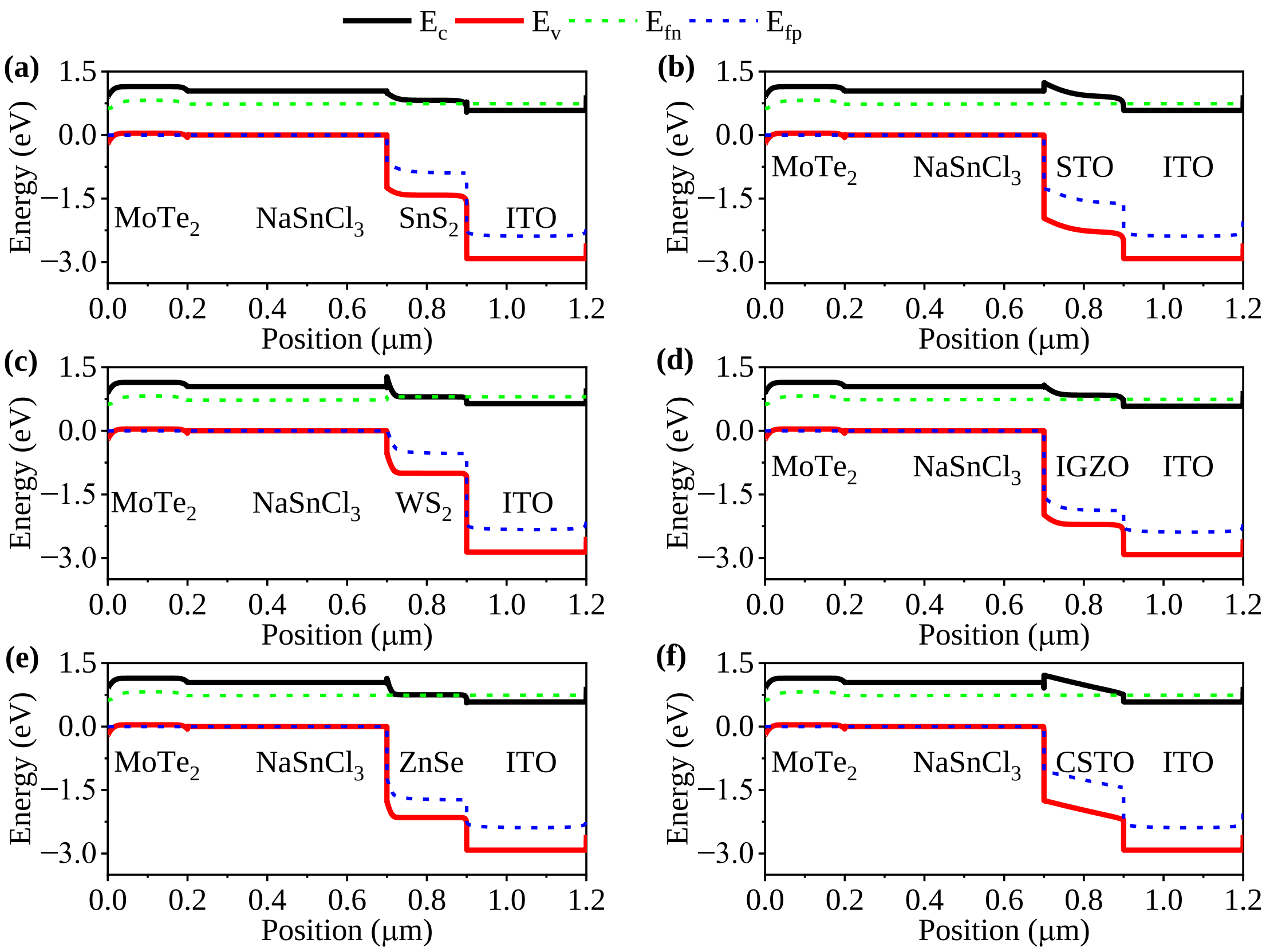

Fig. 6. Energy band diagram of the structures with HTL of $MoTe_2$ and ETL of (a) $SnS_2$, (b) STO, (c) $WS_2$, (d) IGZO, (e) ZnSe, and (f) CSTO. $E_c$ and $E_v$ represent conduction and valence band edge respectively, while $E_{fn}$ and $E_{fp}$ correspond to quasi-Fermi energy level for electron and hole respectively.

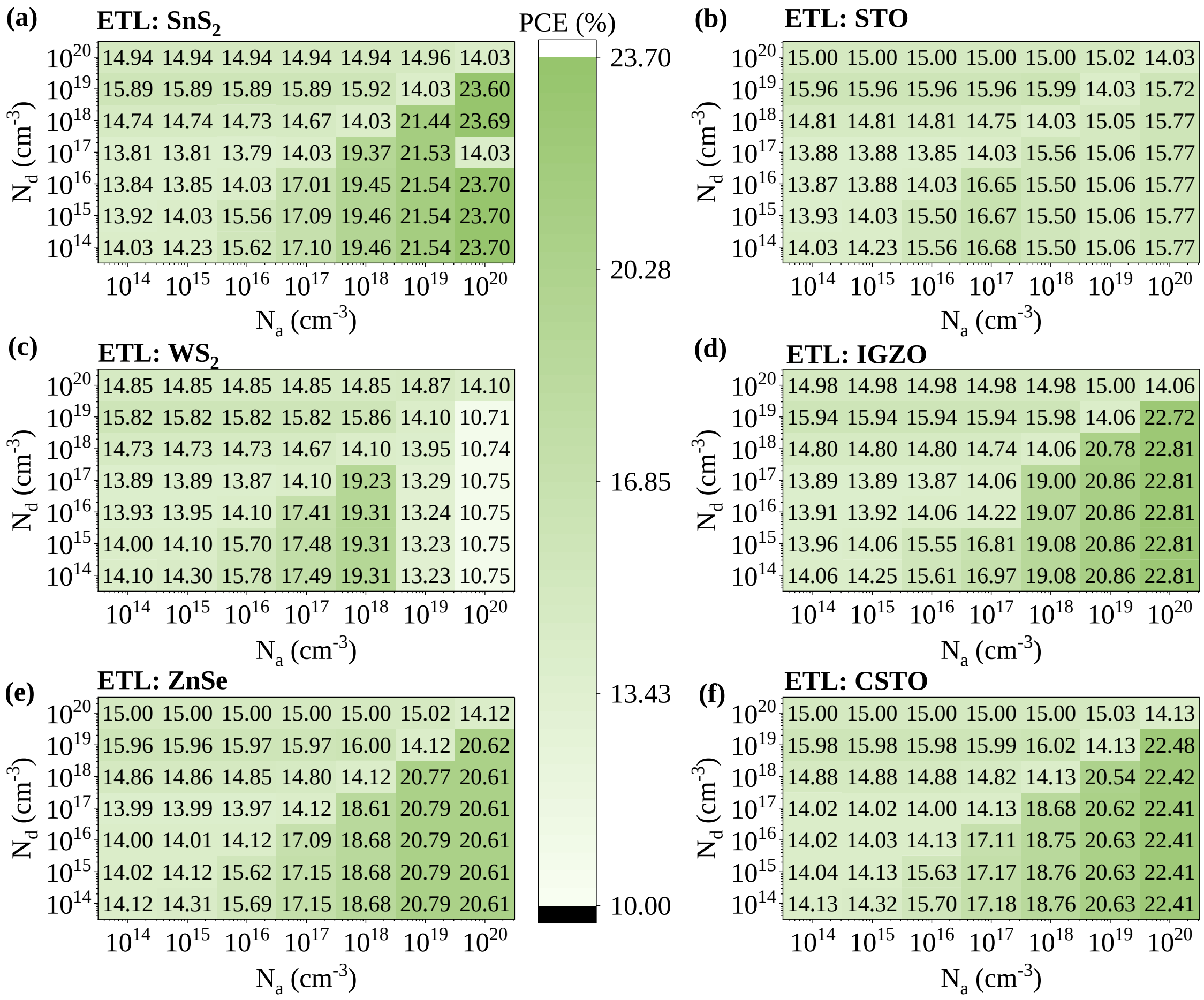

Fig. 7. Effect of donor density, $N_d$ and acceptor density, $N_a$ of $NaSnCl_3$ on PCE of the structures with HTL of $MoTe_2$ and ETL of (a) $SnS_2$, (b) STO, (c) $WS_2$, (d) IGZO, (e) ZnSe, and (f) CSTO.

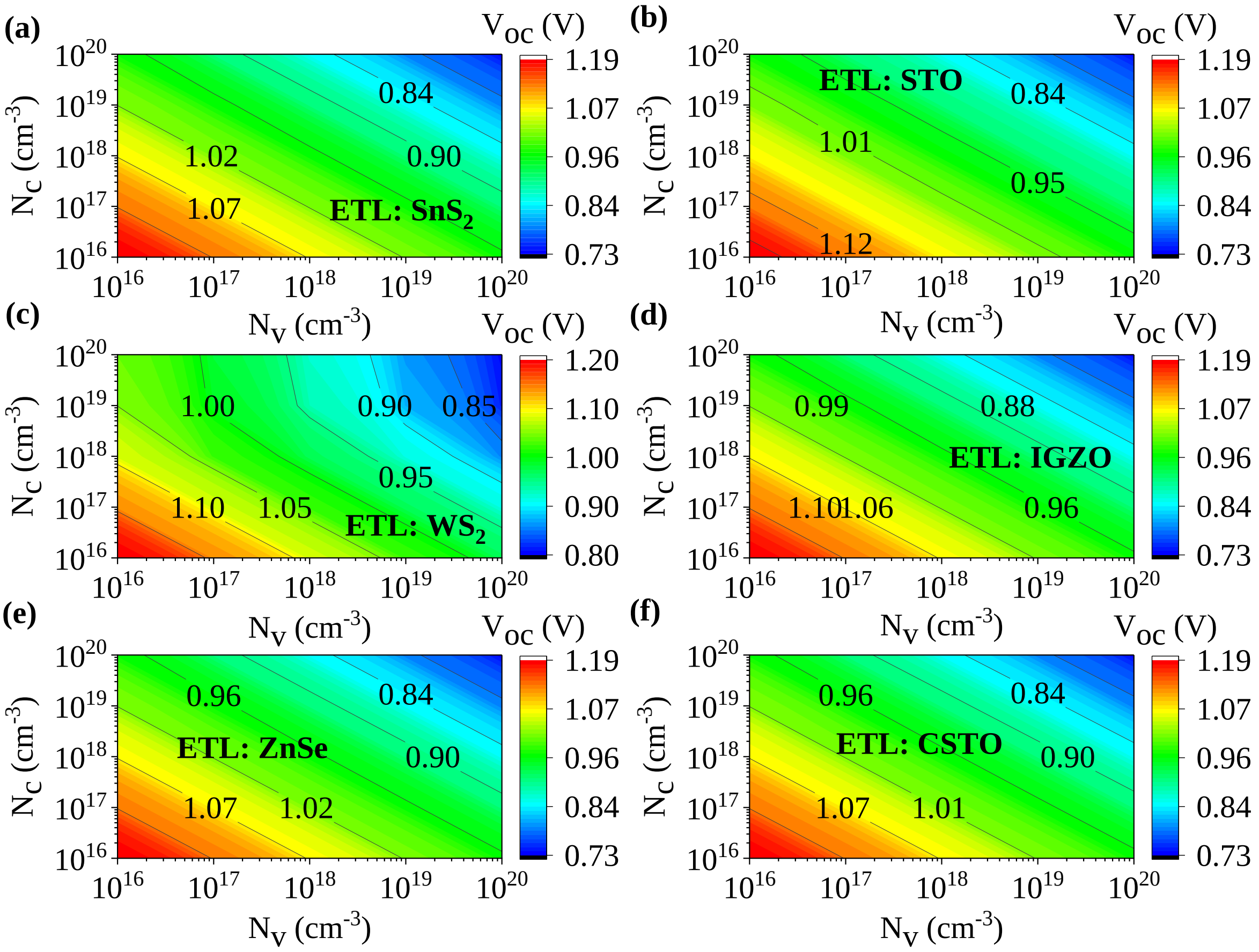

Fig. 8. Impact of effective density of states in conduction and valence band ($N_c$ and $N_v$ respectively) of $NaSnCl_3$ on $V_{oc}$ of the structures with HTL of $MoTe_2$ and ETL of (a) $SnS_2$, (b) STO, (c) $WS_2$, (d) IGZO, (e) ZnSe, and (f) CSTO.

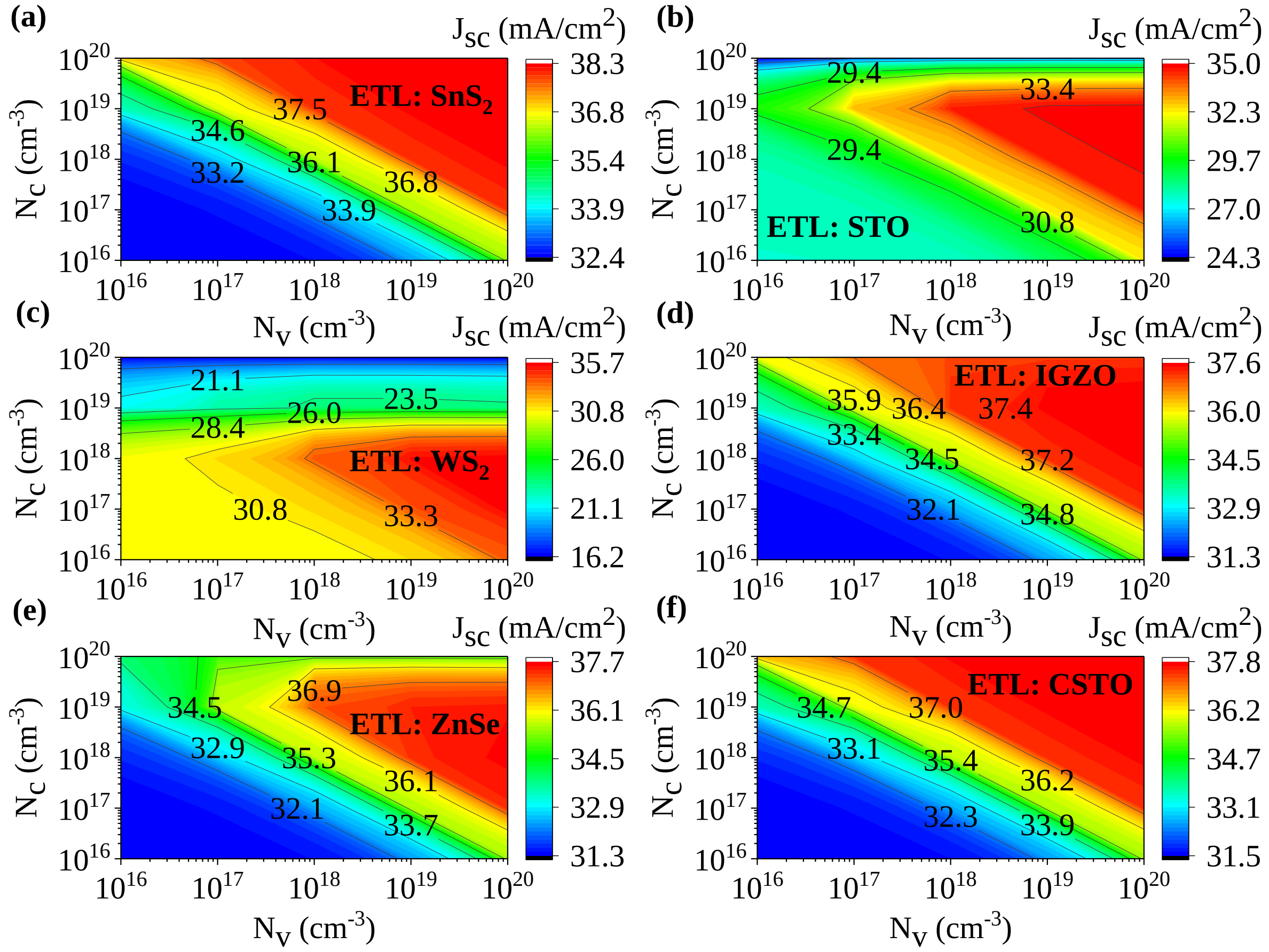

Fig. 9. Impact of effective density of states in conduction and valence band ($N_c$ and $N_v$ respectively) of $NaSnCl_3$ on $J_{sc}$ of the structures with HTL of $MoTe_2$ and ETL of (a) $SnS_2$, (b) STO, (c) $WS_2$, (d) IGZO, (e) ZnSe, and (f) CSTO.

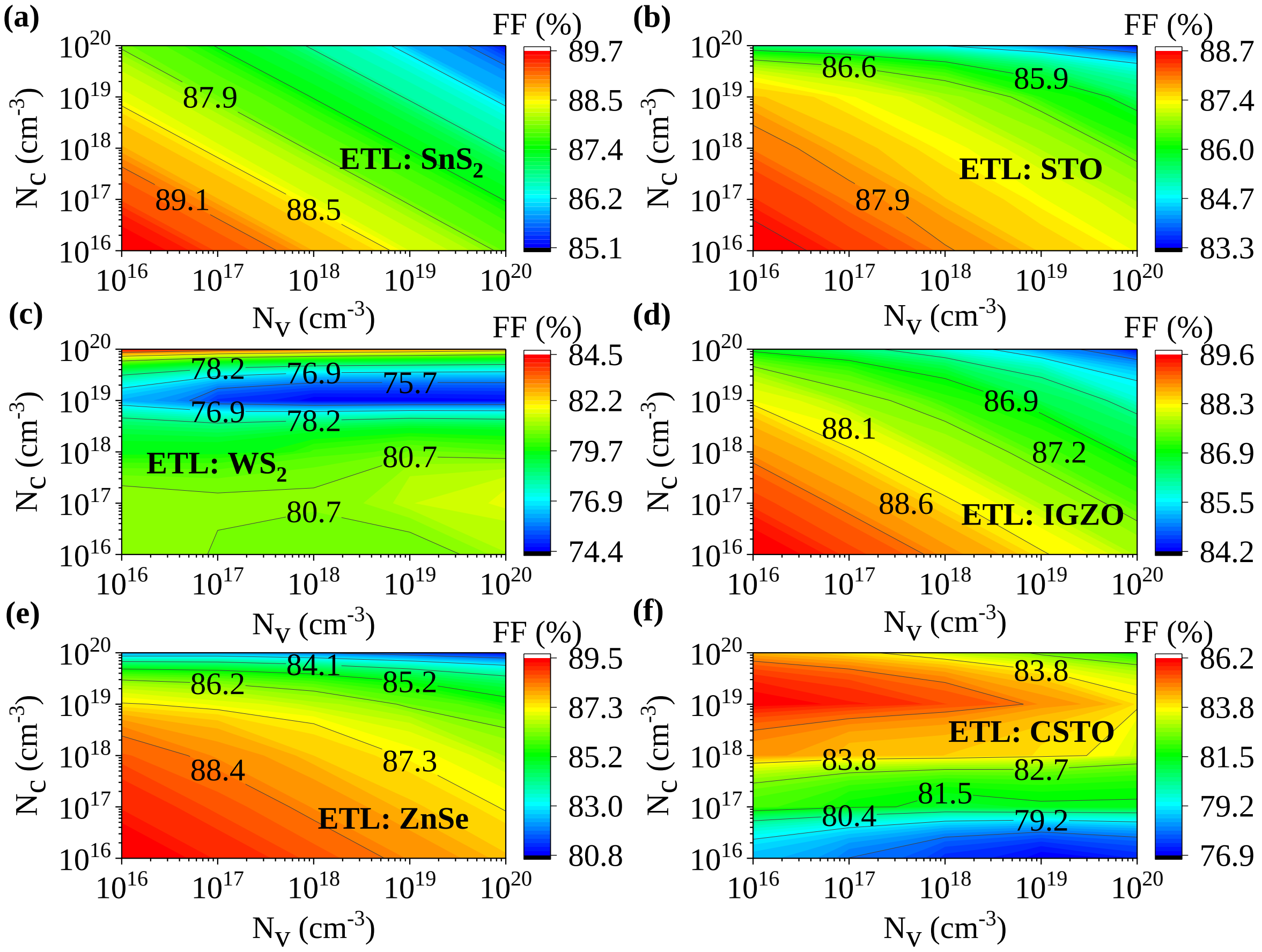

Fig. 10. Impact of effective density of states in conduction and valence band ($N_c$ and $N_v$ respectively) of $NaSnCl_3$ on FF of the structures with HTL of $MoTe_2$ and ETL of (a) $SnS_2$, (b) STO, (c) $WS_2$, (d) IGZO, (e) ZnSe, and (f) CSTO.

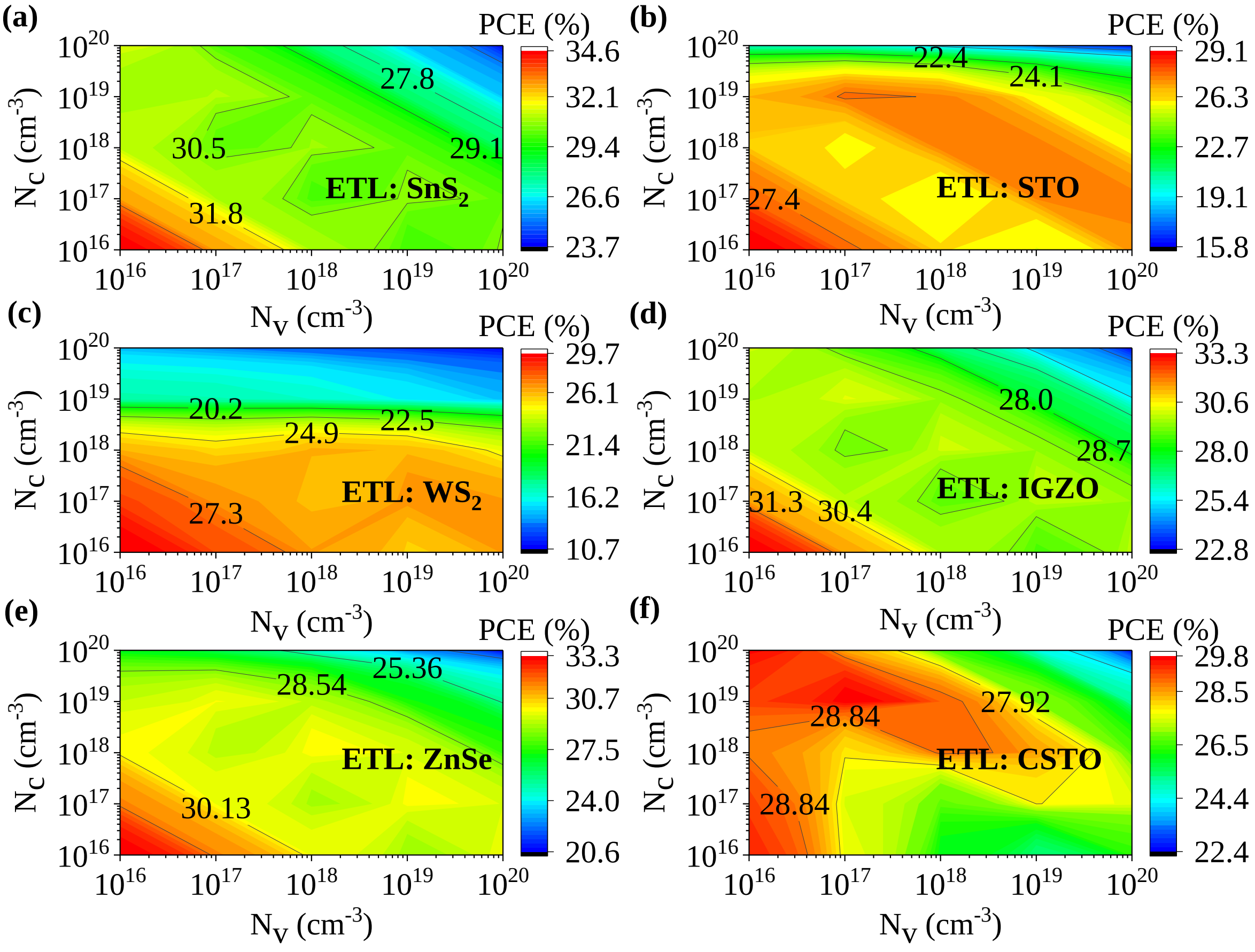

Fig. 11. Impact of effective density of states in conduction and valence band ($N_c$ and $N_v$ respectively) of $NaSnCl_3$ on PCE of the structures with HTL of $MoTe_2$ and ETL of (a) $SnS_2$, (b) STO, (c) $WS_2$, (d) IGZO, (e) ZnSe, and (f) CSTO.

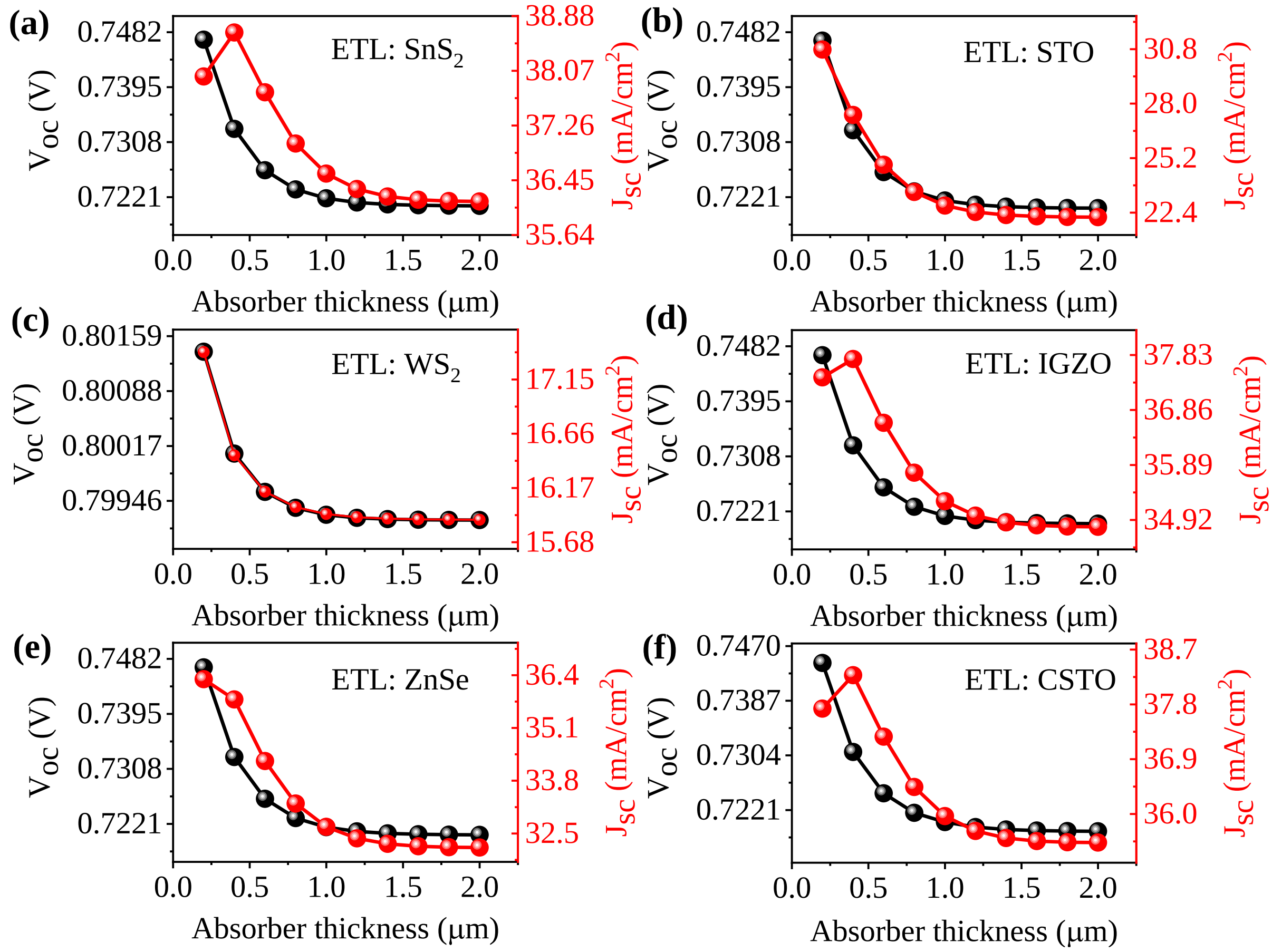

Fig. 12. Effect of thickness of $NaSnCl_3$ on $V_{oc}$ and $J_{sc}$ of the structures with HTL of $MoTe_2$ and ETL of (a) $SnS_2$, (b) STO, (c) $WS_2$, (d) IGZO, (e) ZnSe, and (f) CSTO.

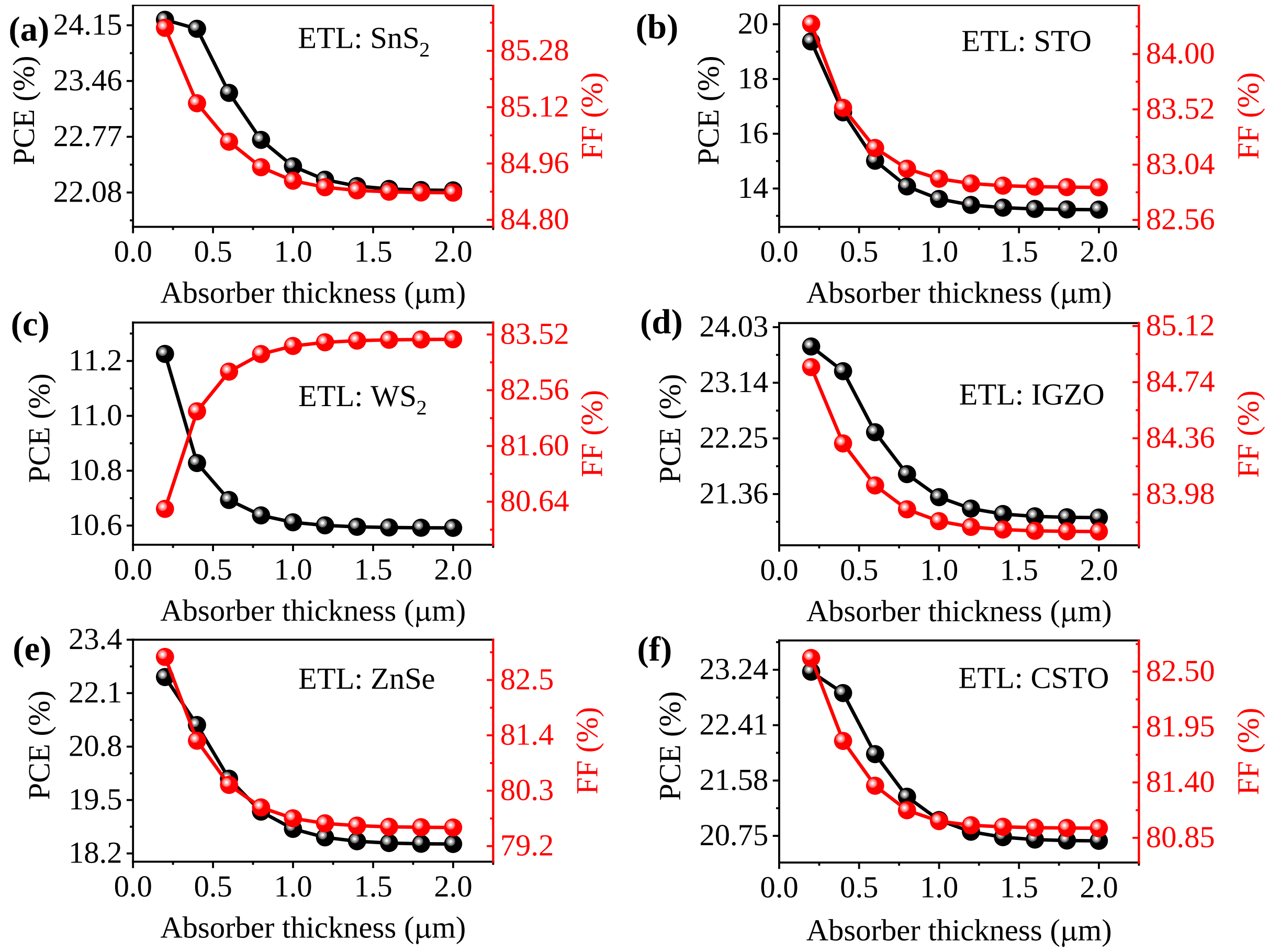

Fig. 13. Effect of thickness of $NaSnCl_3$ on PCE and FF of the structures with HTL of $MoTe_2$ and ETL of (a) $SnS_2$, (b) STO, (c) $WS_2$, (d) IGZO, (e) ZnSe, and (f) CSTO.

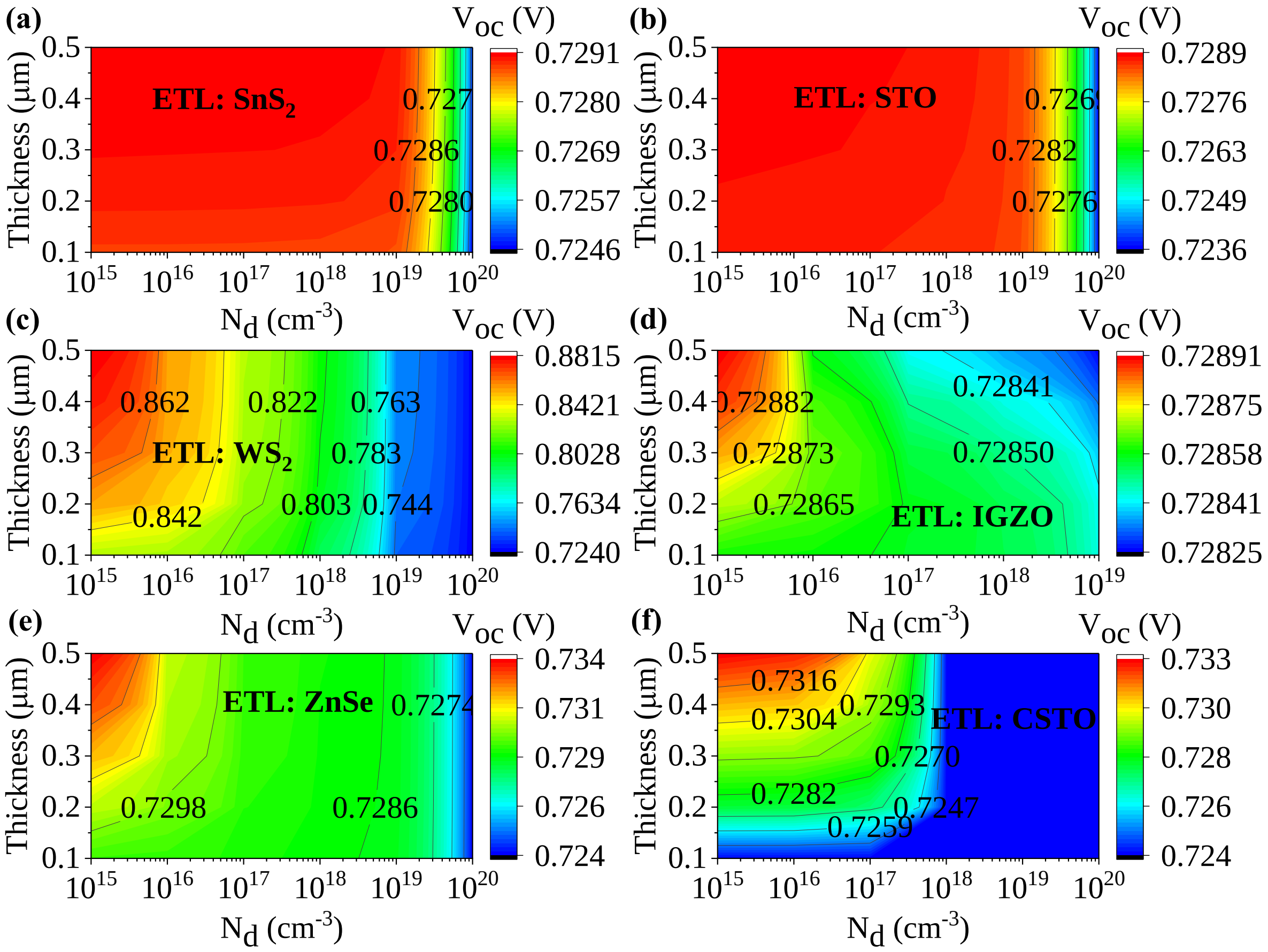

Fig. 14. Effect of thickness and doping density of ETL material, $N_d$ on $V_{oc}$ of the structures with HTL of $MoTe_2$ and ETL of (a) $SnS_2$, (b) STO, (c) $WS_2$, (d) IGZO, (e) ZnSe, and (f) CSTO.

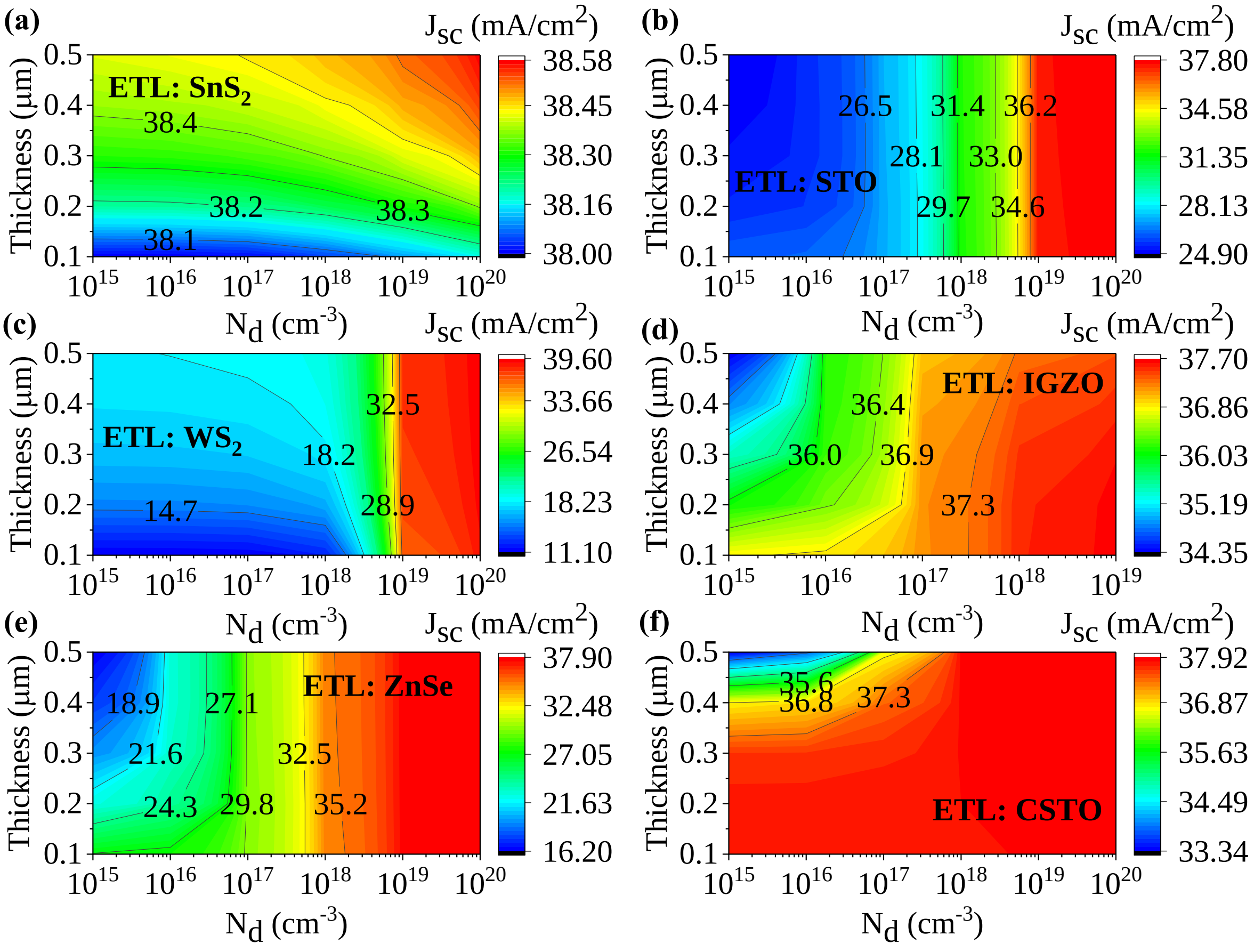

Fig. 15. Effect of thickness and doping density of ETL material, $N_d$ on $J_{sc}$ of the structures with HTL of $MoTe_2$ and ETL of (a) $SnS_2$, (b) STO, (c) $WS_2$, (d) IGZO, (e) ZnSe, and (f) CSTO.

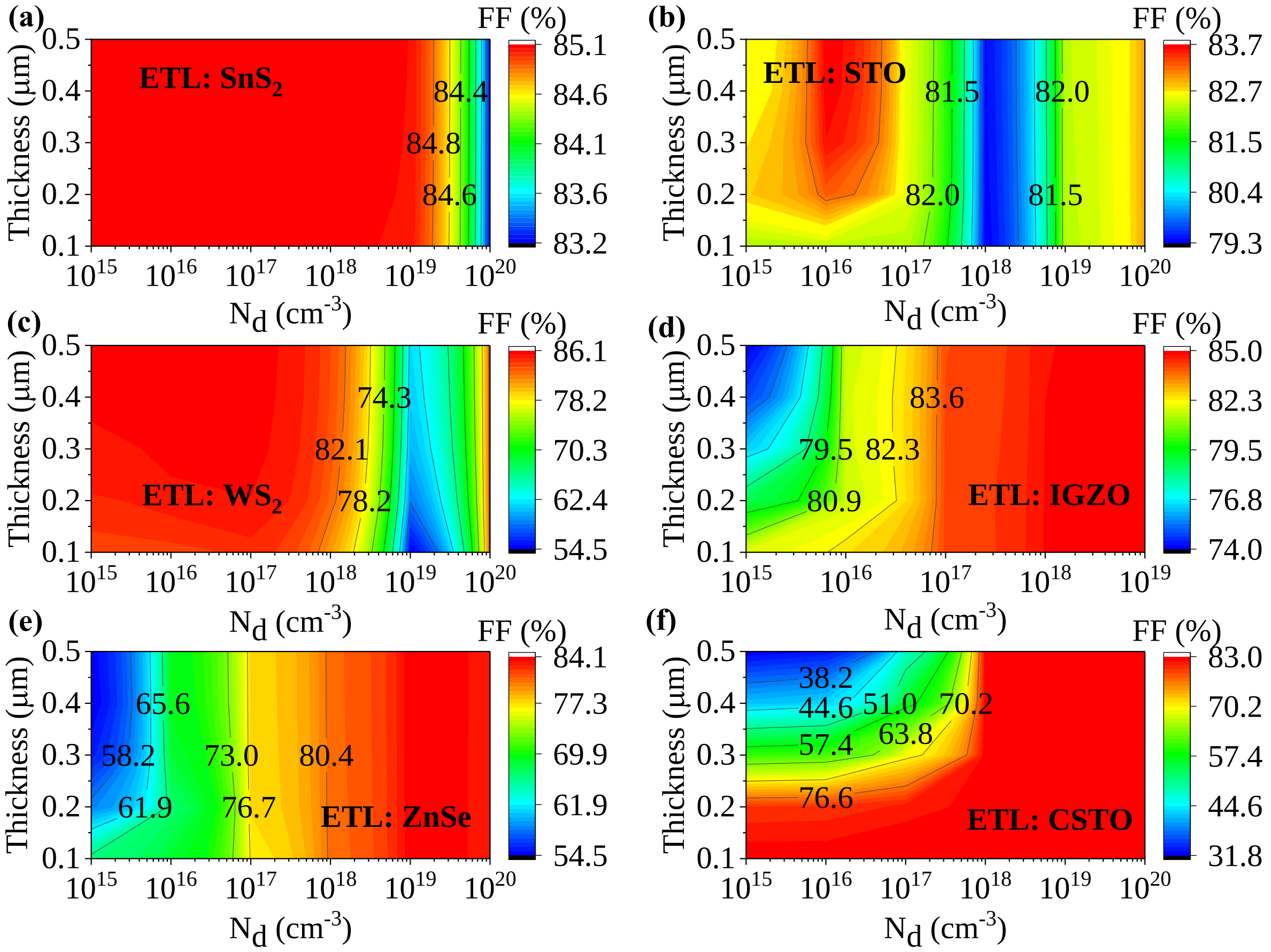

Fig. 16. Effect of thickness and doping density of ETL material, $N_d$ on FF of the structures with HTL of $MoTe_2$ and ETL of (a) $SnS_2$, (b) STO, (c) $WS_2$, (d) IGZO, (e) ZnSe, and (f) CSTO.

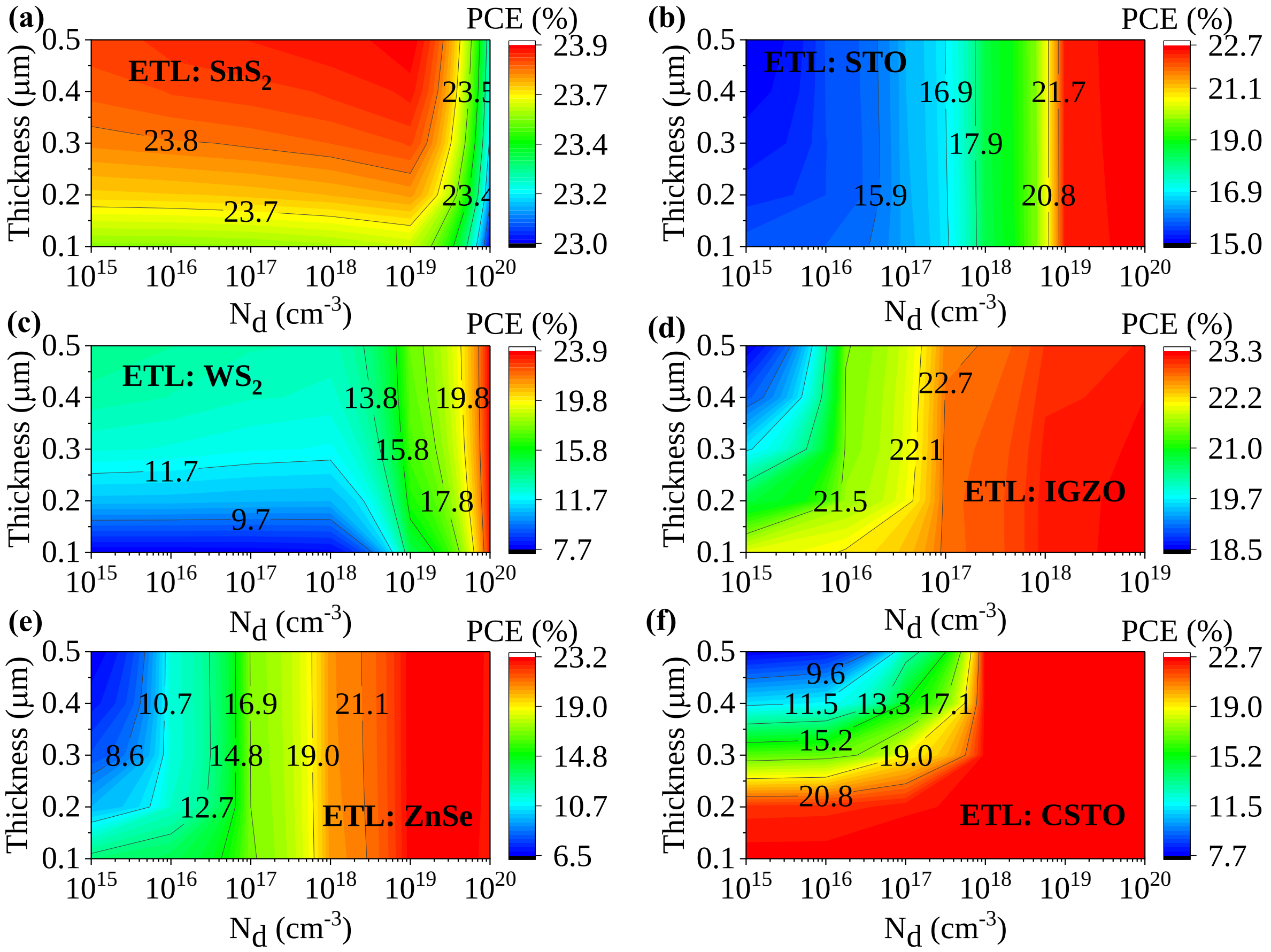

Fig. 17. Effect of thickness and doping density of ETL material, $N_d$ on PCE of the structures with HTL of $MoTe_2$ and ETL of (a) $SnS_2$, (b) STO, (c) $WS_2$, (d) IGZO, (e) ZnSe, and (f) CSTO.

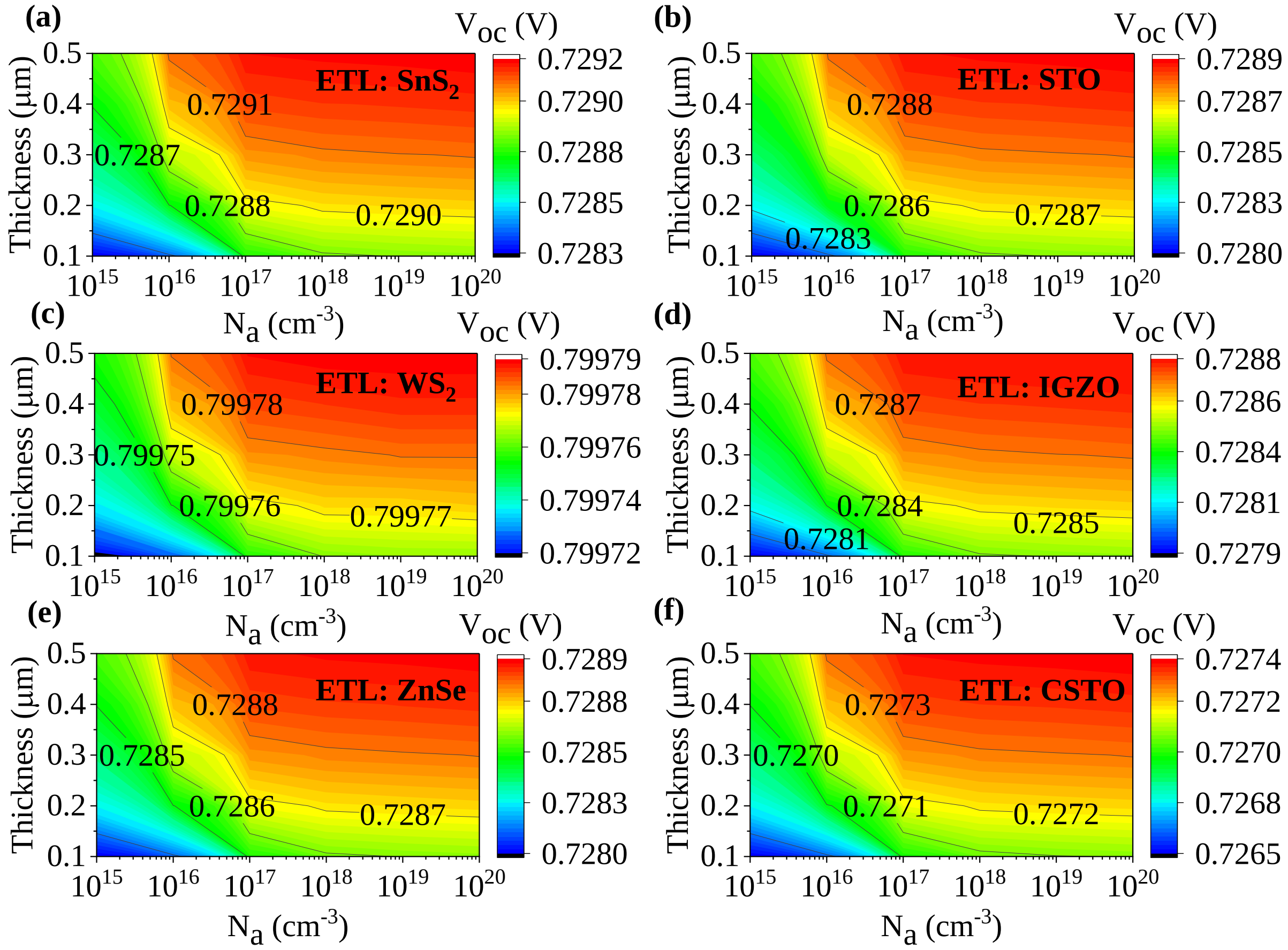

Fig. 18. Effect of thickness and doping density of HTL material, $N_a$ on $V_{oc}$ of the structures with HTL of $MoTe_2$ and ETL of (a) $SnS_2$, (b) STO, (c) $WS_2$, (d) IGZO, (e) ZnSe, and (f) CSTO.

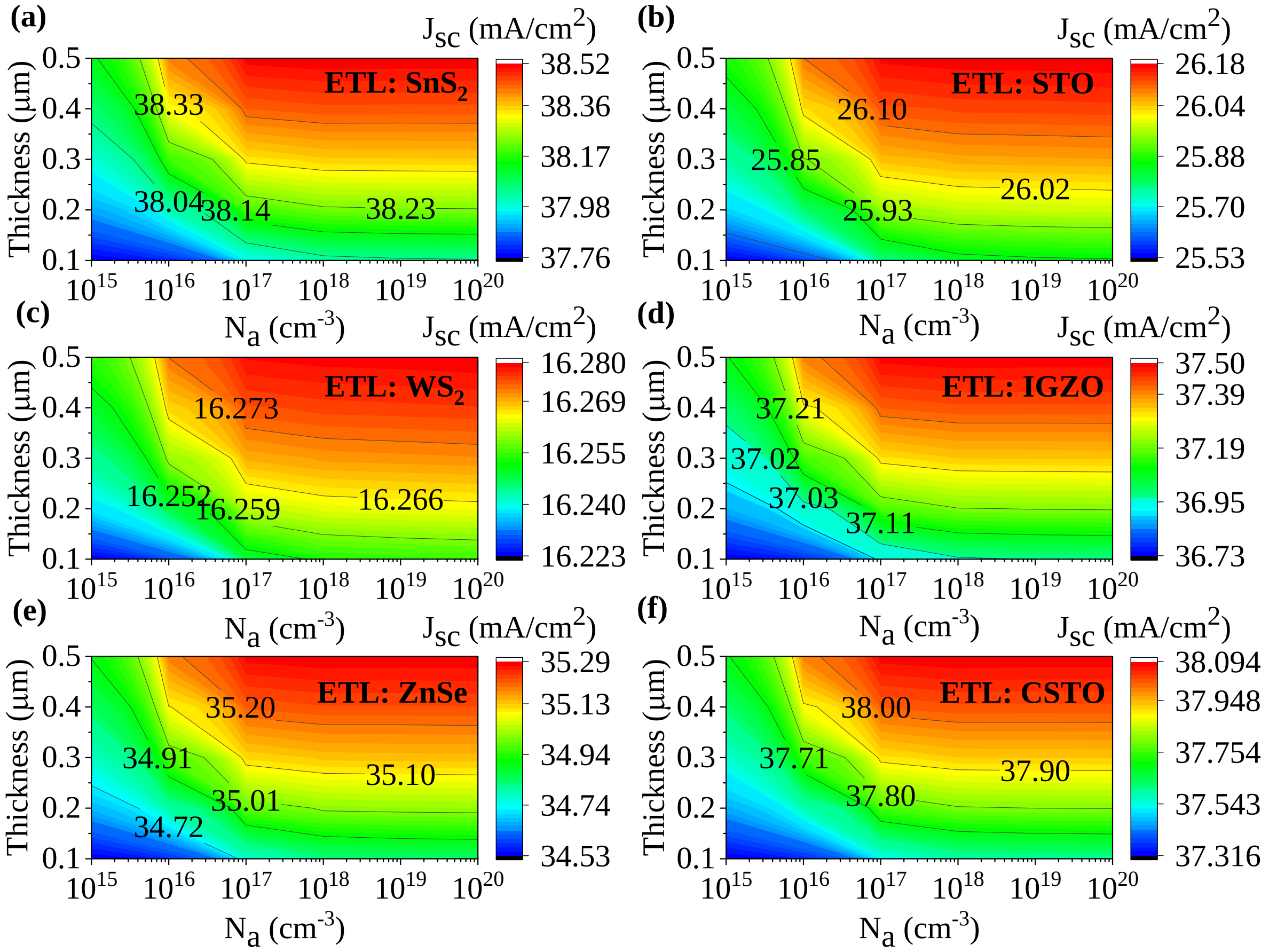

Fig. 19. Effect of thickness and doping density of HTL material, $N_a$ on $J_{sc}$ of the structures with HTL of $MoTe_2$ and ETL of (a) $SnS_2$, (b) STO, (c) $WS_2$, (d) IGZO, (e) ZnSe, and (f) CSTO.

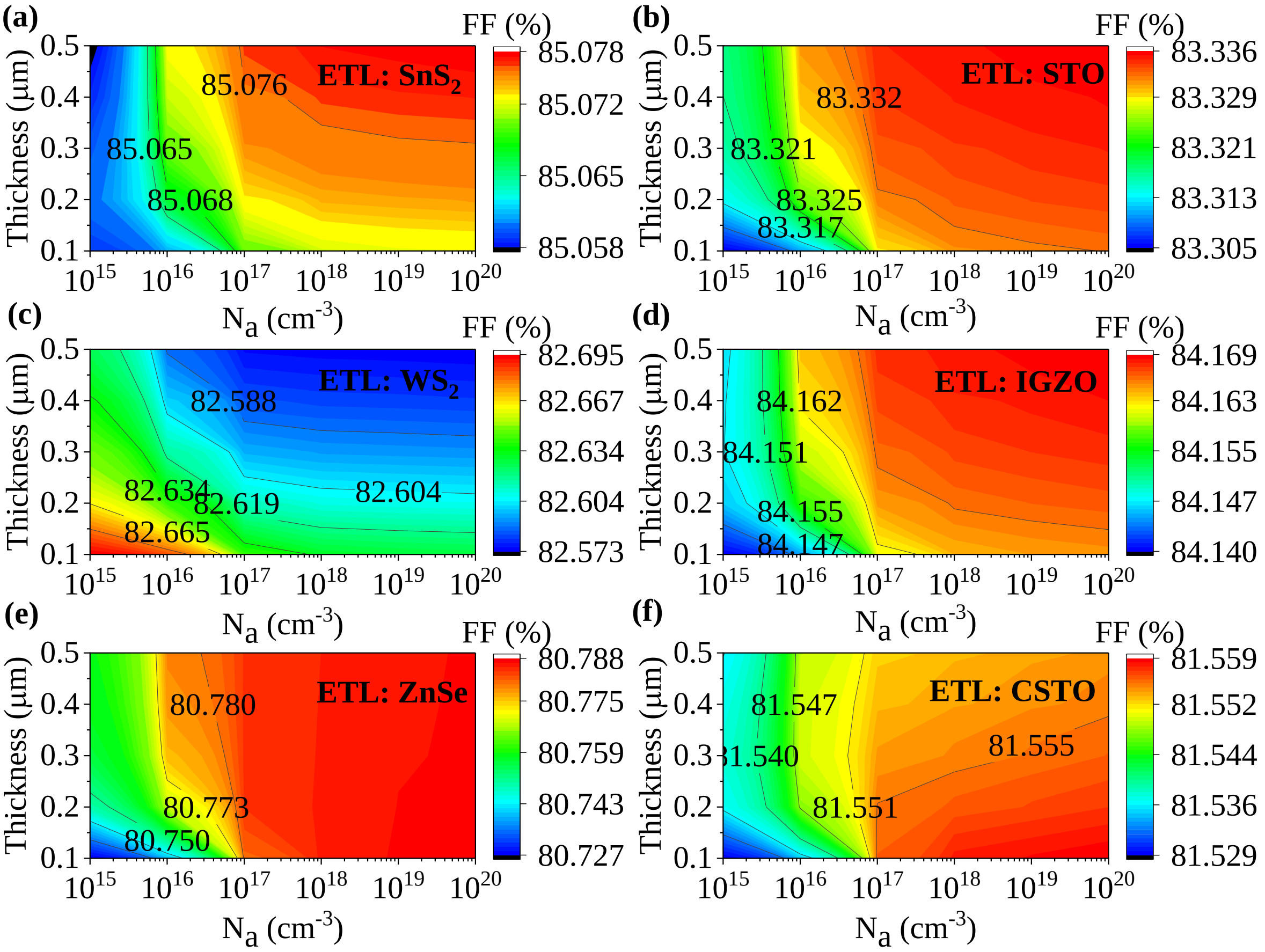

Fig. 20. Effect of thickness and doping density of HTL material, $N_a$ on FF of the structures with HTL of $MoTe_2$ and ETL of (a) $SnS_2$, (b) STO, (c) $WS_2$, (d) IGZO, (e) ZnSe, and (f) CSTO.

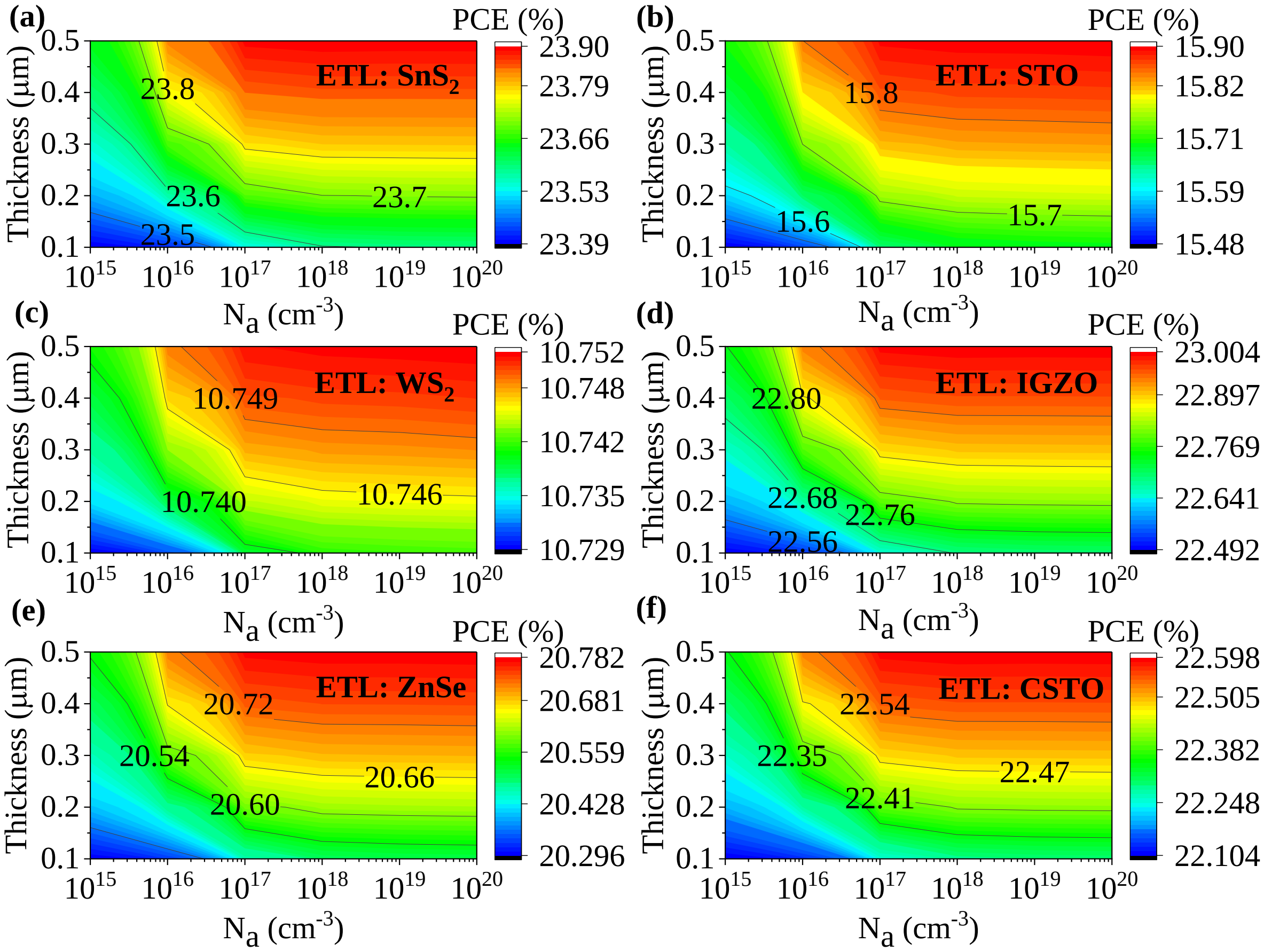

Fig. 21. Effect of thickness and doping density of HTL material, $N_a$ on PCE of the structures with HTL of $MoTe_2$ and ETL of (a) $SnS_2$, (b) STO, (c) $WS_2$, (d) IGZO, (e) ZnSe, and (f) CSTO.

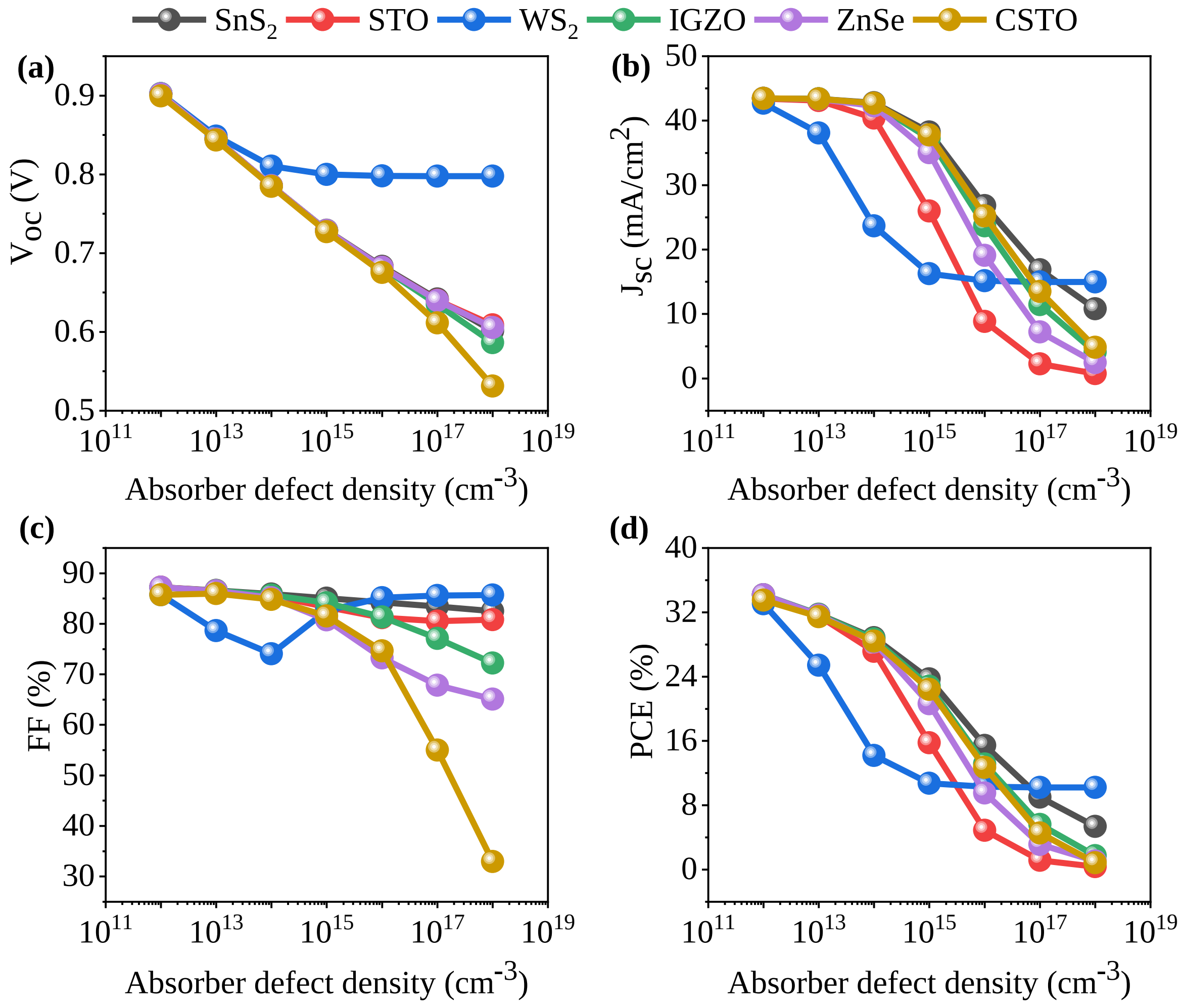

Fig. 22. Effect of defects in $NaSnCl_3$ on PV performance parameters: (a) $V_{oc}$, (b) $J_{sc}$, (c) FF, and (d) PCE of the structures with $MoTe_2$ as HTL and $SnS_2$, STO, $WS_2$, IGZO, ZnSe, and CSTO as ETLs.

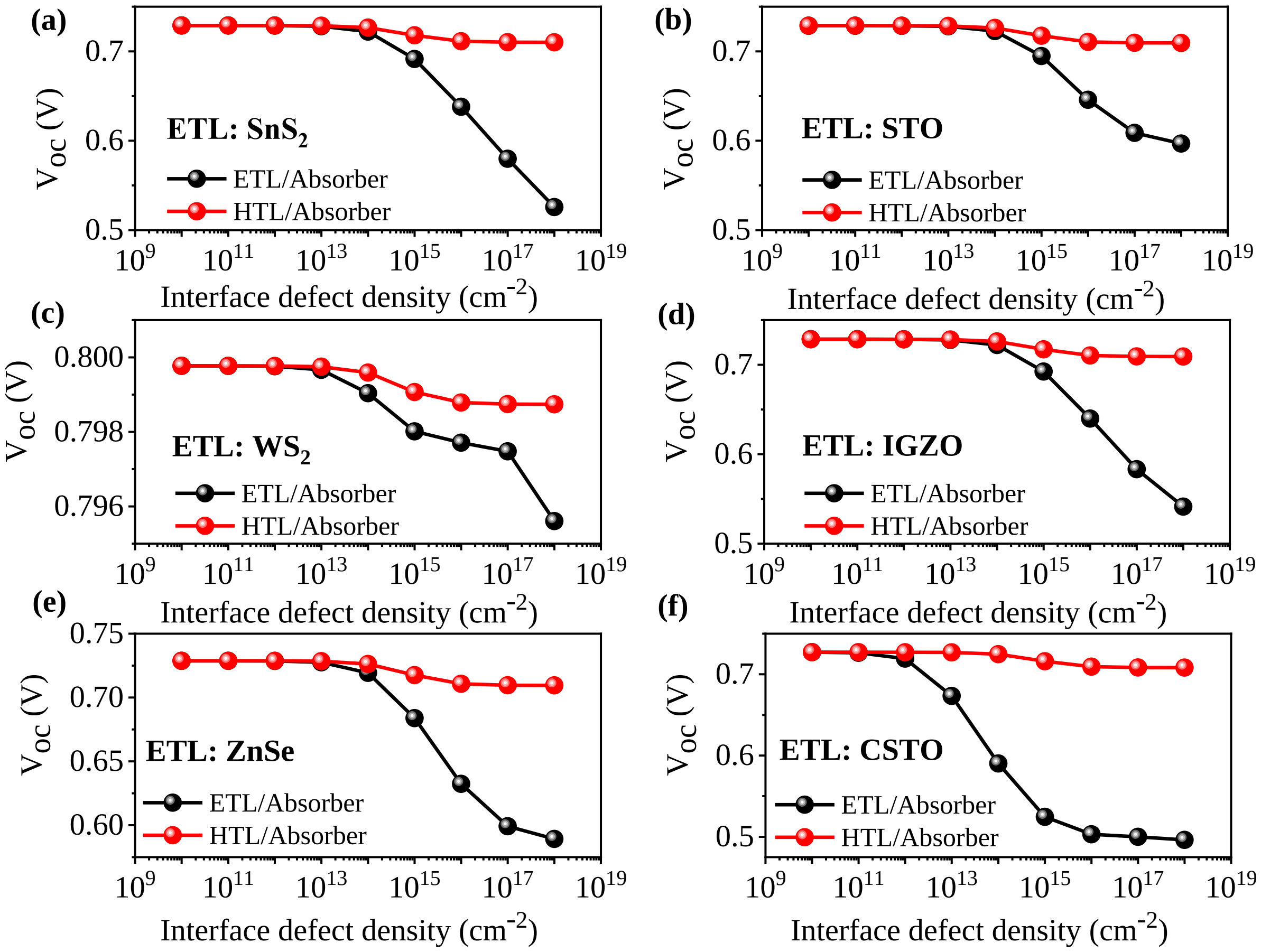

Fig. 23. Effect of defects at ETL/$NaSnCl_3$ and HTL/$NaSnCl_3$ interfaces on $V_{oc}$ of the structures with $MoTe_2$ as HTL and (a) $SnS_2$, (b) STO, (c) $WS_2$, (d) IGZO, (e) ZnSe, and (f) CSTO as ETL.

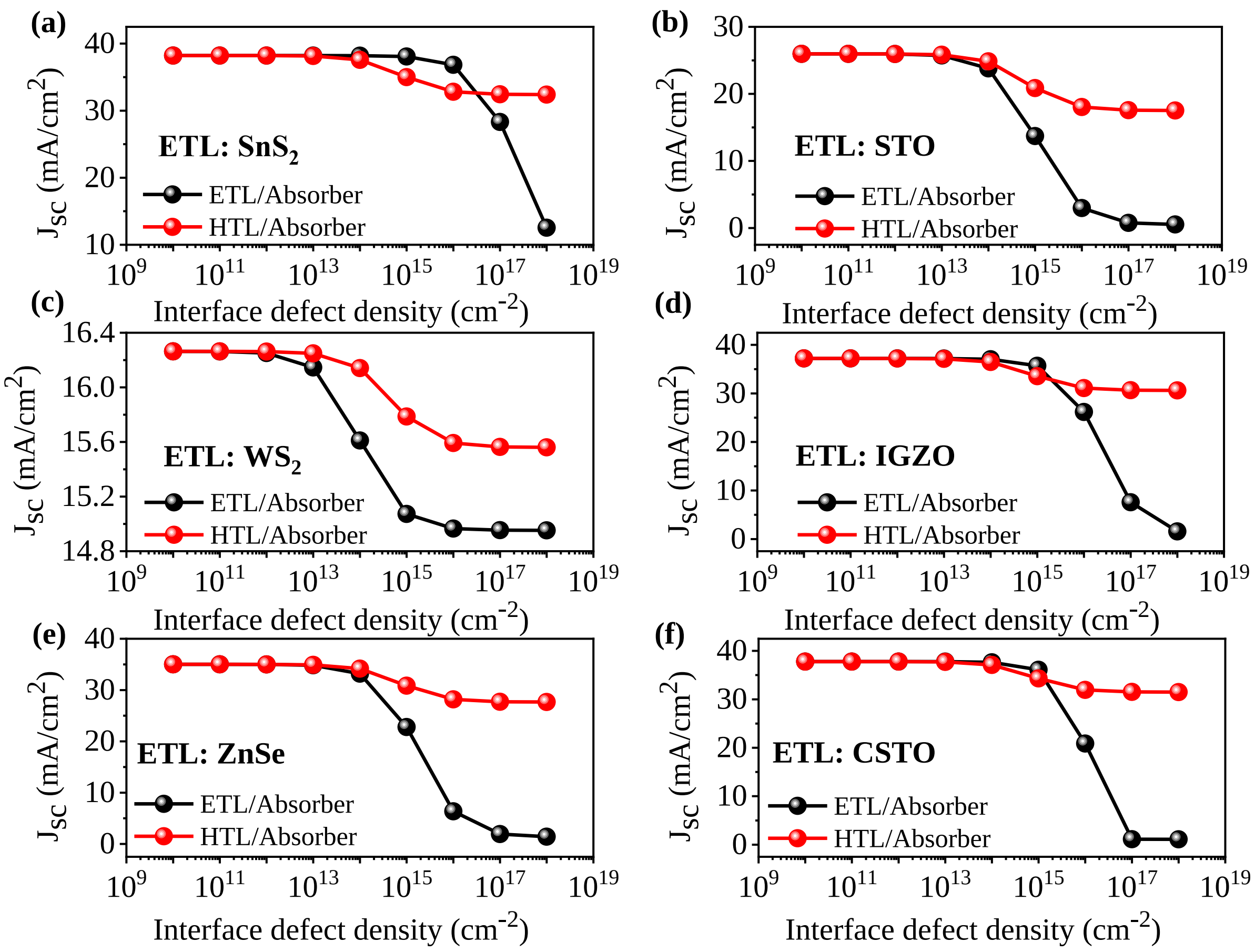

Fig. 24. Effect of defects at ETL/$NaSnCl_3$ and HTL/$NaSnCl_3$ interfaces on $J_{sc}$ of the structures with $MoTe_2$ as HTL and (a) $SnS_2$, (b) STO, (c) $WS_2$, (d) IGZO, (e) ZnSe, and (f) CSTO as ETL.

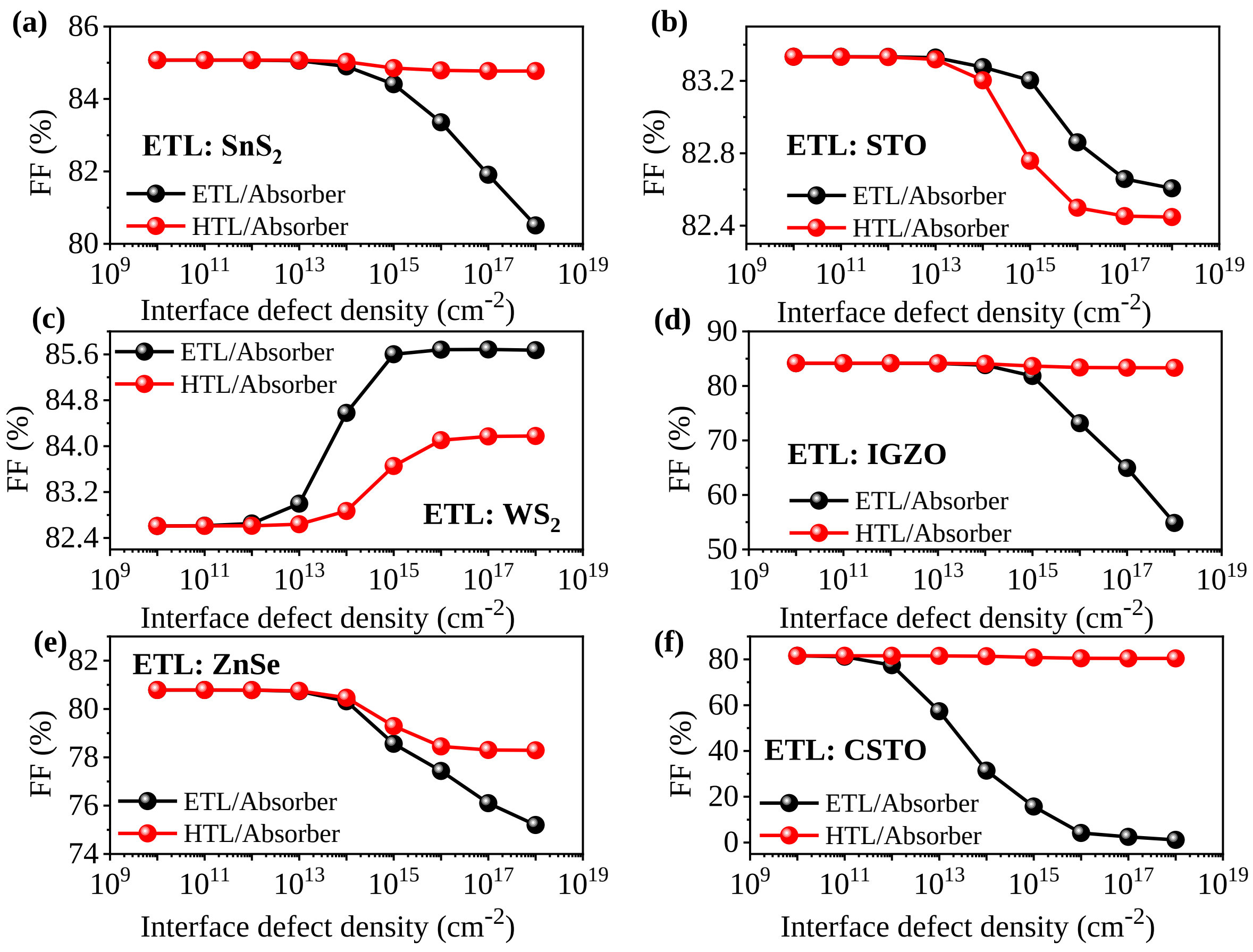

Fig. 25. Effect of defects at ETL/$NaSnCl_3$ and HTL/$NaSnCl_3$ interfaces on FF of the structures with $MoTe_2$ as HTL and (a) $SnS_2$, (b) STO, (c) $WS_2$, (d) IGZO, (e) ZnSe, and (f) CSTO as ETL.

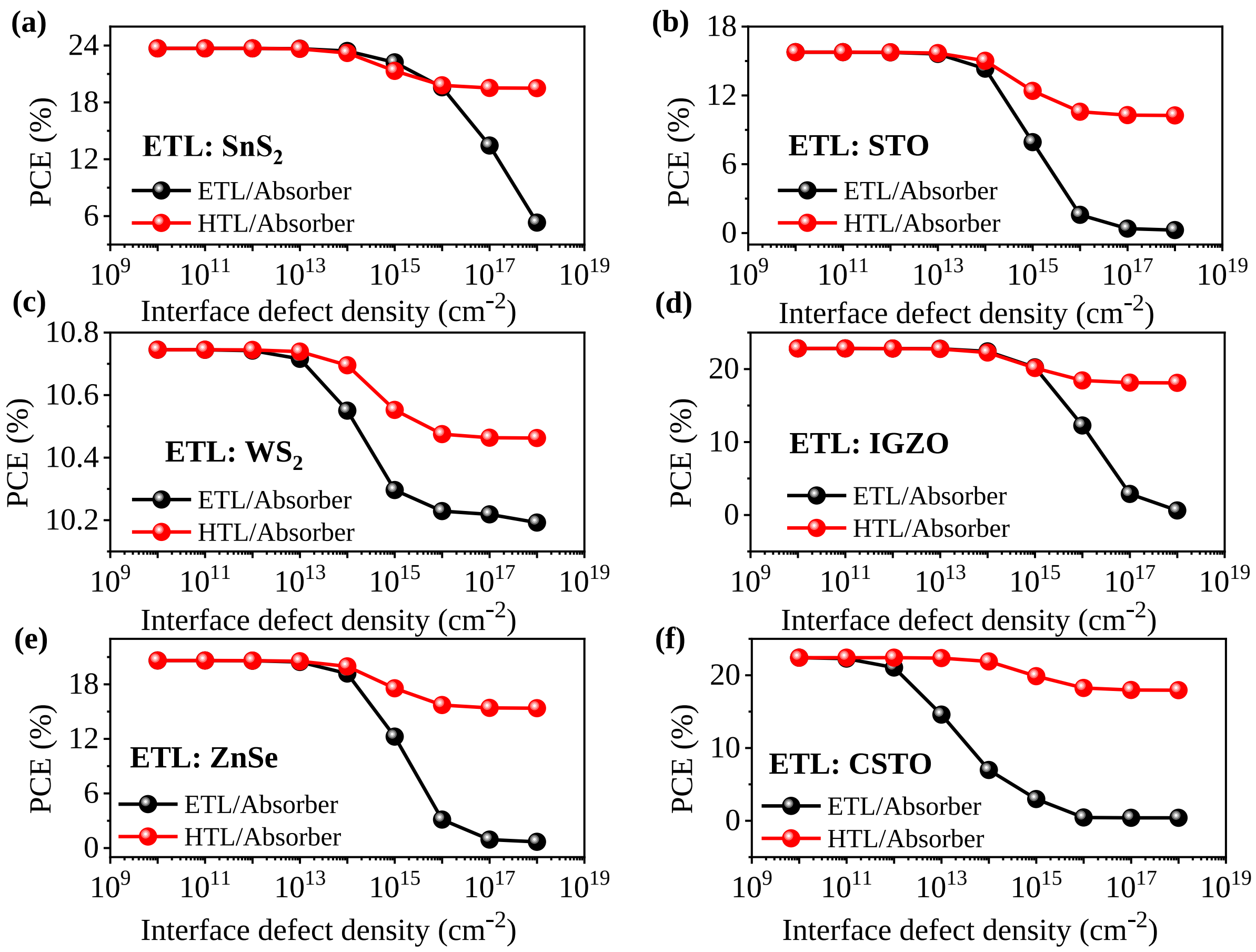

Fig. 26. Effect of defects at ETL/$NaSnCl_3$ and HTL/$NaSnCl_3$ interfaces on PCE of the structures with $MoTe_2$ as HTL and (a) $SnS_2$, (b) STO, (c) $WS_2$, (d) IGZO, (e) ZnSe, and (f) CSTO as ETL.

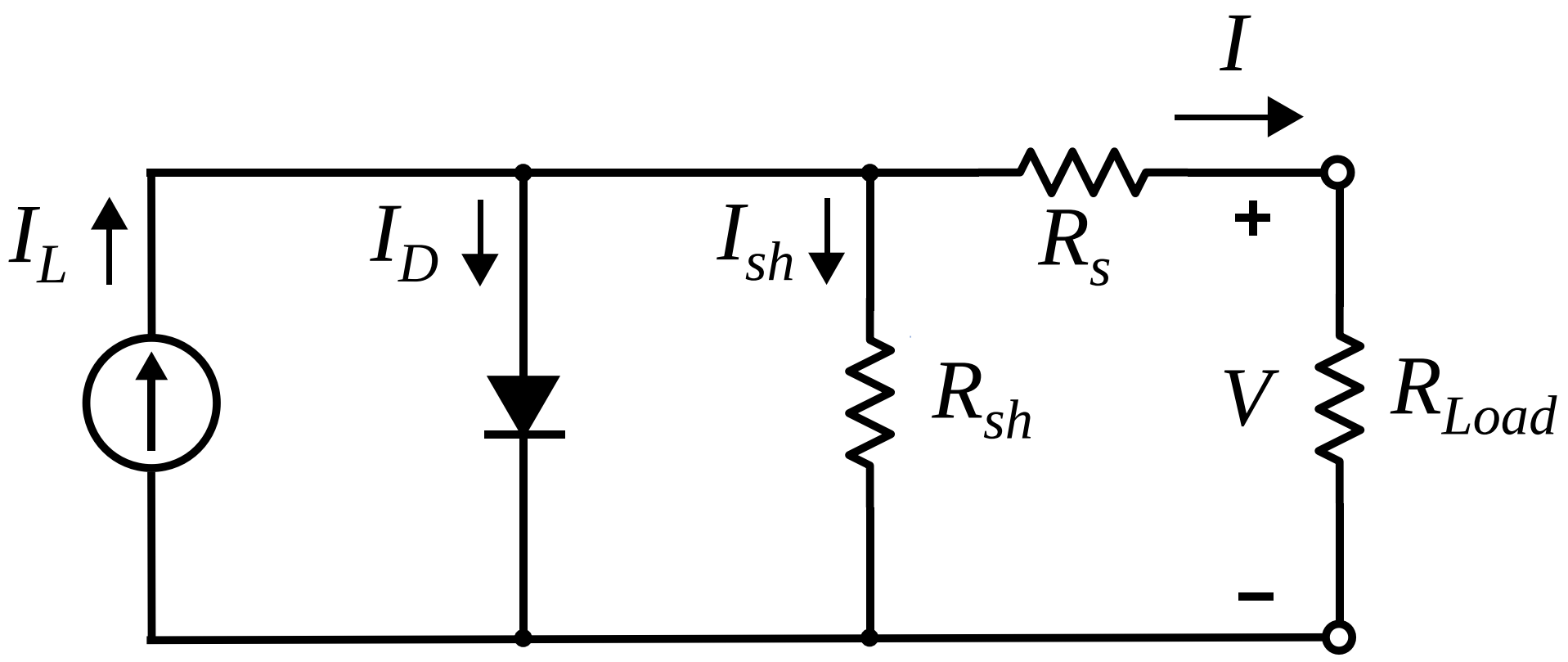

Fig. 27. Equivalent circuit model of a solar cell. I is the current delivered to the load, $I_L$ is the photocurrent, $I_D$ is the dark current, $I_{sh}$ is the current through the shunt resistance, V is the voltage across the load. $R_s$ and $R_{sh}$ are series and shunt resistance respectively.

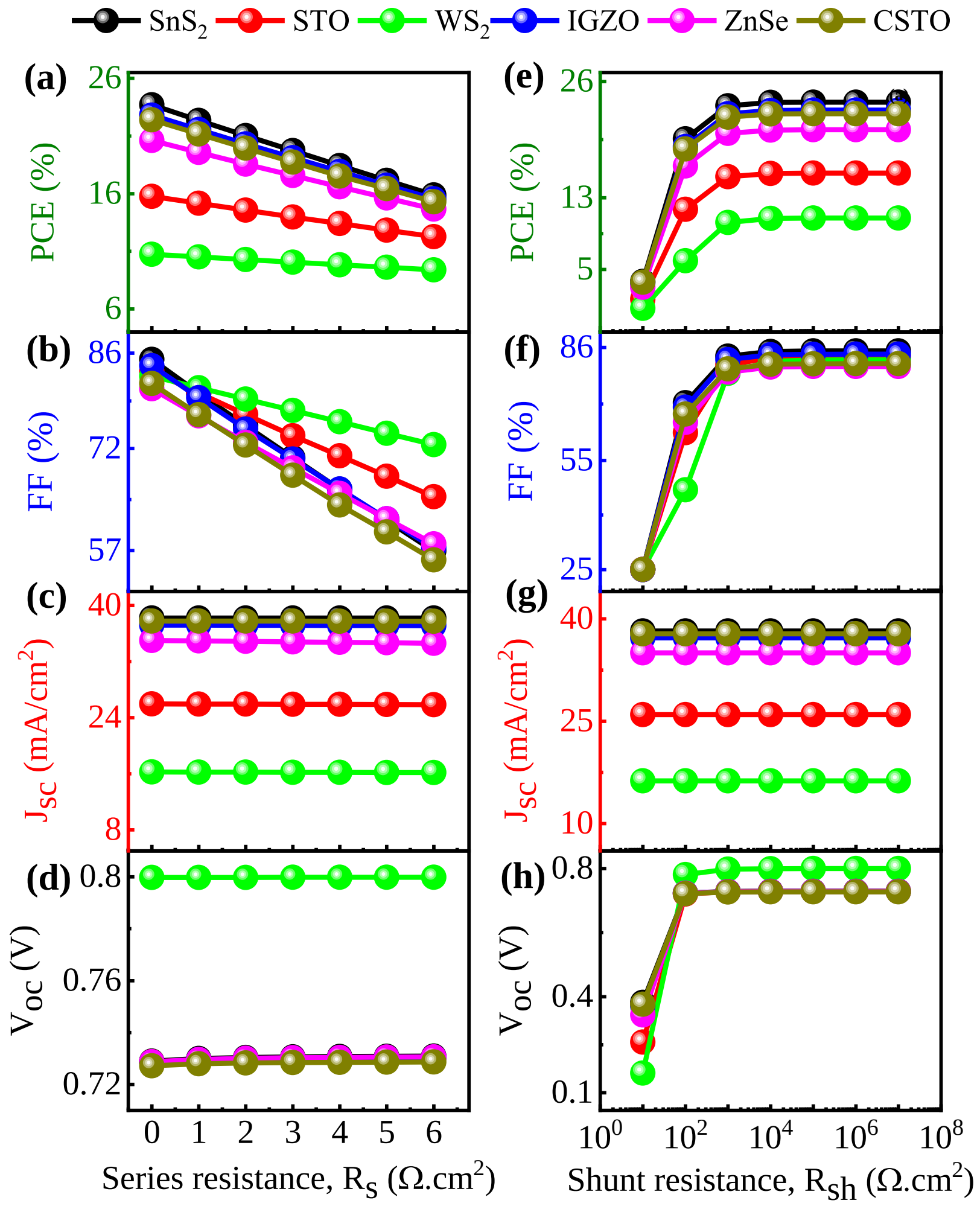

Fig. 28. Effect of series and shunt resistance on PV performance parameters: (a) and (e) PCE, (b) and (f) FF, (c) and (g) $J_{sc}$, and (d) and (h) $V_{oc}$ of the structures with $MoTe_2$ as HTL and $SnS_2$, STO, $WS_2$, IGZO, ZnSe, and CSTO as ETLs.

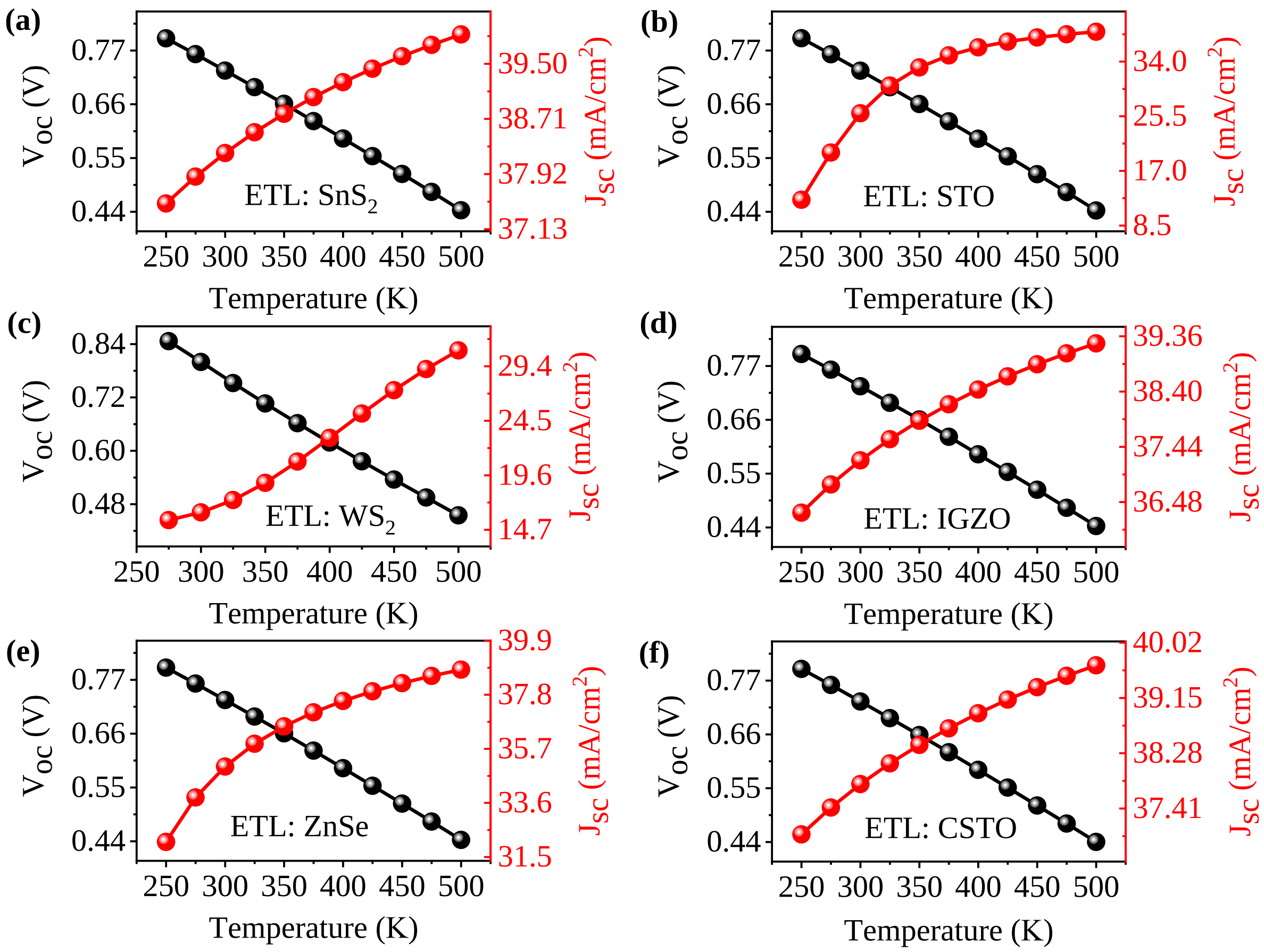

Fig. 29. Effect of temperature on $V_{oc}$ and $J_{sc}$ of the structures with HTL of $MoTe_2$ and ETL of (a) $SnS_2$, (b) STO, (c) $WS_2$, (d) IGZO, (e) ZnSe, and (f) CSTO.

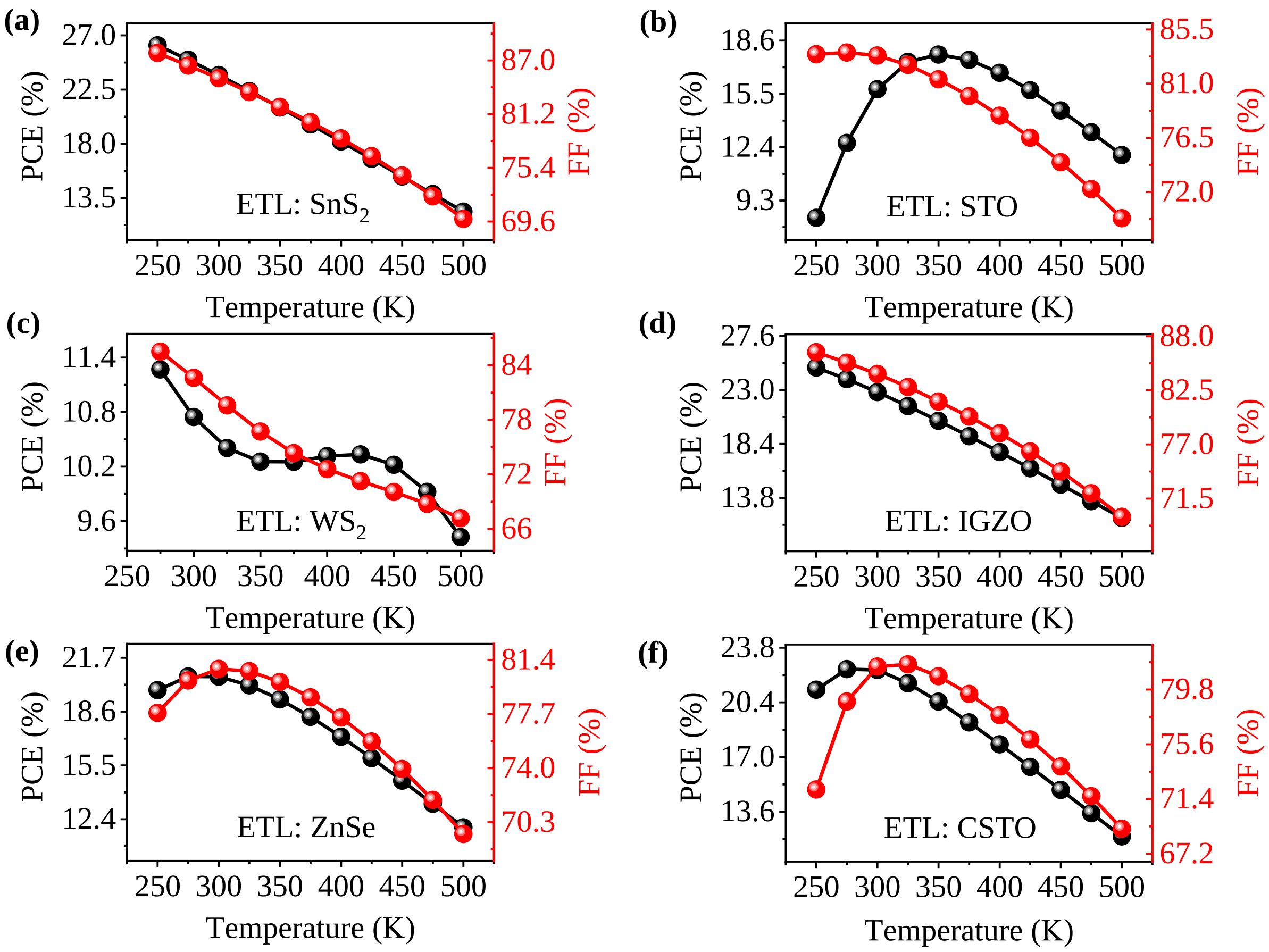

Fig. 30. Effect of temperature on PCE and FF of the structures with HTL of $MoTe_2$ and ETL of (a) $SnS_2$, (b) STO, (c) $WS_2$, (d) IGZO, (e) ZnSe, and (f) CSTO.

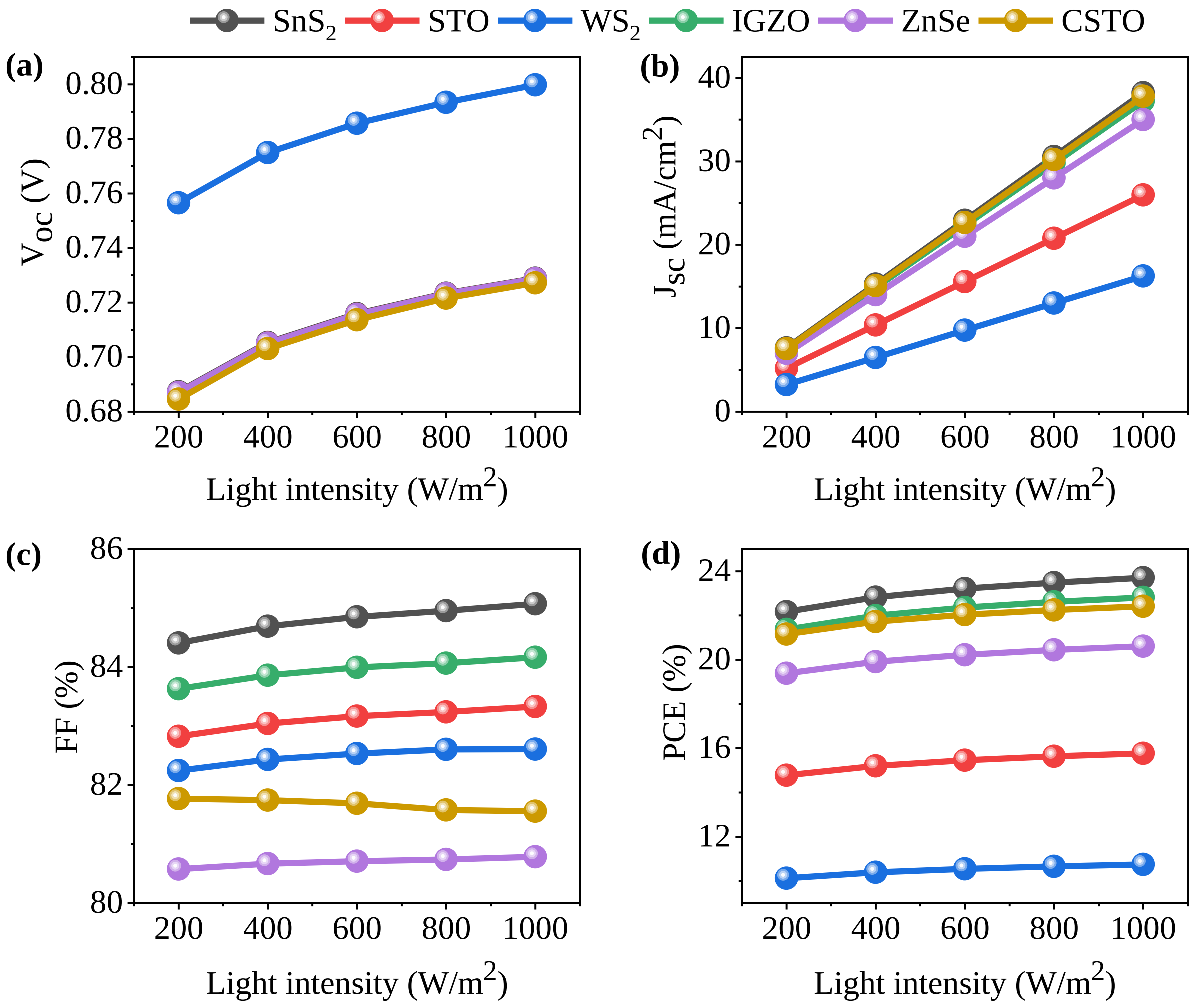

Fig. 31. Impact of incident light intensity on PV performance parameters: (a) $V_{oc}$, (b) $J_{sc}$, (c) FF, and (d) PCE of the structures with $MoTe_2$ as HTL and $SnS_2$, STO, $WS_2$, IGZO, ZnSe, and CSTO as ETLs.

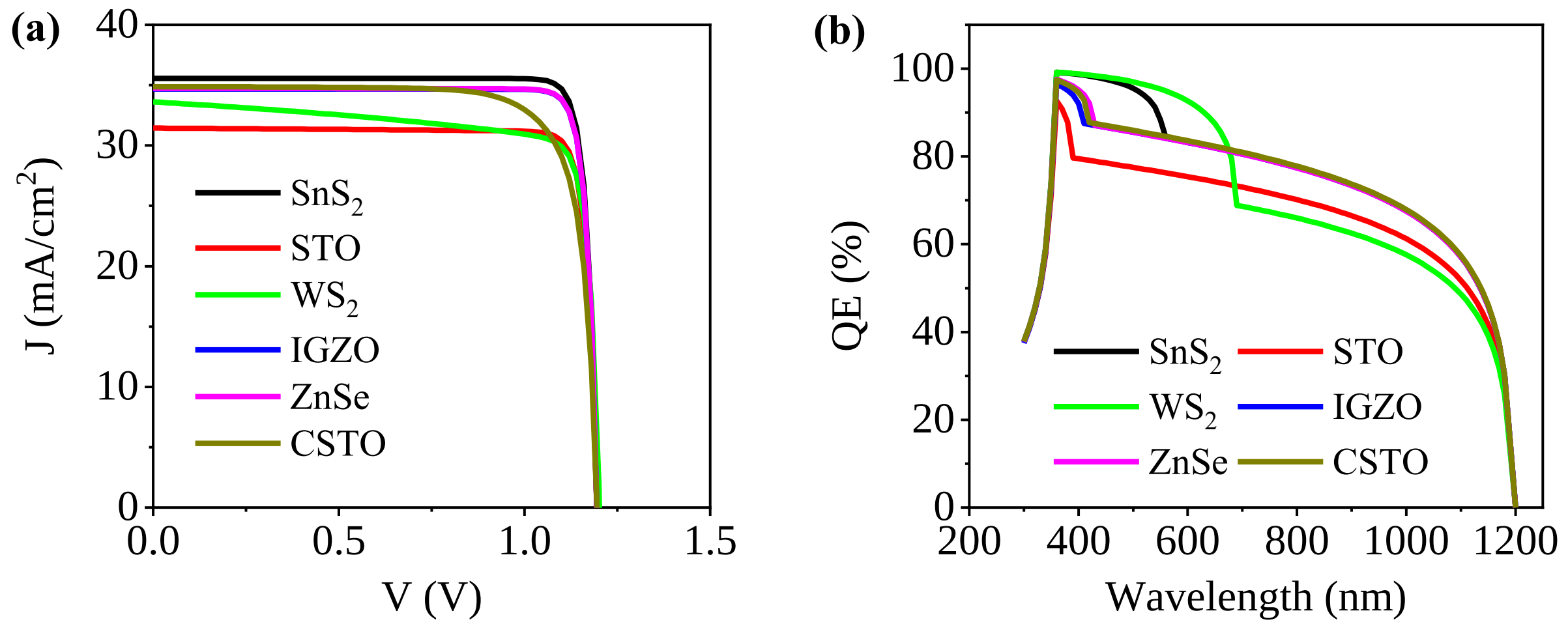

Fig. 32. (a) Current density versus voltage and (b) quantum efficiency versus wavelength of light for six optimized structures with $MoTe_2$ as HTL and $SnS_2$, STO, $WS_2$, IGZO, ZnSe and CSTO as ETL.

Table 1. Input parameters of ITO, $NaSnCl_3$, and HTLs

| Parameters | ITO | $NaSnCl_3$ | CuSCN | GO | Mg–$CuCrO_2$ | Spiro-OMeTAD | CdTe | GaAs | $MoTe_2$ | $BaSi_2$ | P3HT |
|---|---|---|---|---|---|---|---|---|---|---|---|
| Thickness (nm) | 300 | 500 | 200 | 200 | 200 | 200 | 200 | 200 | 200 | 200 | 200 |
| Band gap, $E_g$ (eV) | 3.5 | 1.04 | 3.6 | 2.48 | 3 | 3 | 1.5 | 1.42 | 1.1 | 1.3 | 1.7 |
| Electron affinity, $\chi$ (eV) | 4 | 4.2 | 1.7 | 2.3 | 2.1 | 2.2 | 3.9 | 4.07 | 4.2 | 3.3 | 3.5 |
| Dielectric permittivity (relative), $\varepsilon_r$ | 9 | 10 | 10 | 10 | 9.5 | 3 | 9.4 | 12.9 | 13 | 11.17 | 3 |
| Conduction band effective density of states, $N_c$ (1/cm$^3$) | $2.2 \times 10^{18}$ | $10^{20}$ | $2.2 \times 10^{19}$ | $2.2 \times 10^{18}$ | $10^{19}$ | $2.2 \times 10^{19}$ | $8 \times 10^{17}$ | $2.2 \times 10^{18}$ | $10^{15}$ | $2.6 \times 10^{19}$ | $2 \times 10^{21}$ |
| Valence band effective density of | $1.8 \times 10^{19}$ | $10^{20}$ | $1.8 \times 10^{18}$ | $1.8 \times 10^{19}$ | $10^{19}$ | $1.8 \times 10^{19}$ | $1.8 \times 10^{19}$ | $1.8 \times 10^{19}$ | $10^{17}$ | $2 \times 10^{19}$ | $2 \times 10^{21}$ |

| Parameters | ITO | $NaSnCl_3$ | CuSCN | GO | Mg–$CuCrO_2$ | Spiro-OMeTAD | CdTe | GaAs | $MoTe_2$ | $BaSi_2$ | P3HT |
|---|---|---|---|---|---|---|---|---|---|---|---|
| states, $N_V$ (1/cm$^3$) | | | | | | | | | | | |
| Electron mobility, $\mu_n$ (cm$^2$/Vs) | 20 | 1 | 100 | 26 | 0.1 | $1 \times 10^{-4}$ | 320 | 8500 | 110 | 820 | $1.8 \times 10^{-3}$ |
| Hole mobility, $\mu_h$ (cm$^2$/Vs) | 10 | 1 | 25 | 123 | 2.53 | $1 \times 10^{-4}$ | 40 | 400 | 426 | 100 | $1.86 \times 10^{-2}$ |
| Shallow uniform acceptor density, $N_a$ (1/cm$^3$) | 0 | $10^{15}$ | $10^{18}$ | $2 \times 10^{18}$ | $6.4 \times 10^{15}$ | $2 \times 10^{19}$ | $2 \times 10^{14}$ | $10^{11}$ | $5 \times 10^{17}$ | $5 \times 10^{18}$ | $10^{18}$ |
| Shallow uniform donor density, $N_d$ (1/cm$^3$) | $1 \times 10^{21}$ | $1 \times 10^{15}$ | 0 | 0 | 0 | 0 | 0 | 0 | 0 | 0 | 0 |
| Defect density, $N_t$ (1/cm$^3$) | $10^{14}$ | $10^{15}$ | $10^{14}$ | $10^{14}$ | $10^{14}$ | $10^{14}$ | $10^{14}$ | $10^{14}$ | $10^{14}$ | $10^{14}$ | $10^{14}$ |
| Reference | [57], [91] | [34], [47], [92] | [57], [93] | [69], [94] | [95] | [47], [96] | [97]–[99] | [97], [100] | [97], [101] | [102], [103] | [57], [93] |

Table 2. Input parameters of ETLs

| Parameters | $TiO_2$ | $SnS_2$ | IGZO | ZnSe | CdS | GaSe | $ZnSnN_2$ | $WS_2$ | PCBM | STO | CSTO |
|---|---|---|---|---|---|---|---|---|---|---|---|
| Thickness (nm) | 200 | 200 | 200 | 200 | 200 | 200 | 200 | 200 | 200 | 200 | 200 |
| Band gap, $E_g$ (eV) | 3.2 | 2.24 | 3.05 | 2.9 | 2.4 | 2.2 | 1.5 | 1.8 | 2 | 3.2 | 2.96 |
| Electron affinity, $X$ (eV) | 4.1 | 4.24 | 4.16 | 4.09 | 4.4 | 4.5 | 4.10 | 3.95 | 3.9 | 4 | 3.9 |

| Parameters | $TiO_2$ | $SnS_2$ | IGZO | ZnSe | CdS | GaSe | $ZnSnN_2$ | $WS_2$ | PCBM | STO | CSTO |
|---|---|---|---|---|---|---|---|---|---|---|---|
| Dielectric permittivity (relative), $\varepsilon_r$ | 10 | 10 | 10 | 10 | 9 | 8 | 15 | 13.6 | 3.9 | 8.7 | 802 |
| Conduction band effective density of states, $N_c$ (1/cm$^3$) | $1 \times 10^{19}$ | $2.2 \times 10^{18}$ | $5 \times 10^{18}$ | $1.5 \times 10^{18}$ | $1.8 \times 10^{19}$ | $1.4 \times 10^{18}$ | $1.2 \times 10^{18}$ | $1 \times 10^{18}$ | $2.5 \times 10^{21}$ | $1.7 \times 10^{19}$ | $2.7 \times 10^{19}$ |
| Valence band effective density of states, $N_v$ (1/cm$^3$) | $1 \times 10^{19}$ | $1.8 \times 10^{19}$ | $5 \times 10^{18}$ | $1.8 \times 10^{18}$ | $2.4 \times 10^{18}$ | $1.49 \times 10^{19}$ | $7.8 \times 10^{19}$ | $2.4 \times 10^{19}$ | $2.5 \times 10^{21}$ | $2 \times 10^{20}$ | $3.5 \times 10^{20}$ |
| Electron mobility, $\mu_n$ (cm$^2$/Vs) | 20 | 500 | 15 | 50 | 100 | 250 | 0.5 | 100 | 0.2 | 5300 | 6000 |
| Hole mobility, $\mu_h$ (cm$^2$/Vs) | 10 | 500 | 0.1 | 20 | 25 | 25 | 0.05 | 100 | 0.2 | 660 | 660 |
| Shallow uniform acceptor density, $N_a$ (1/cm$^3$) | 0 | 0 | 0 | 0 | 0 | 0 | 0 | 0 | 0 | 0 | 0 |
| Shallow uniform donor density, $N_d$ (1/cm$^3$) | $10^{15}$ | $10^{17}$ | $10^{17}$ | $10^{18}$ | $10^{15}$ | $10^{17}$ | $10^{19}$ | $10^{18}$ | $2.93 \times 10^{17}$ | $2 \times 10^{16}$ | $10^{17}$ |
| Defect density, $N_t$ (1/cm$^3$) | $10^{14}$ | $10^{14}$ | $10^{14}$ | $10^{14}$ | $10^{14}$ | $10^{14}$ | $10^{14}$ | $10^{14}$ | $10^{14}$ | $10^{14}$ | $10^{14}$ |
| Reference | [47], [96] | [104], [105] | [57], [91], [93] | [106], [107] | [108] | [69], [109] | [69], [110] | [57], [91], [93] | [57], [91], [93] | [111], [112] | [113] |

Table 3. Performance parameters of six solar cell devices with better performance

| Device structure | $V_{oc}$ (V) | $J_{sc}$ (mA/cm$^2$) | FF (%) | PCE (%) |
|---|---|---|---|---|
| Al/ITO/$SnS_2$/$NaSnCl_3$/$MoTe_2$/Au | 0.4663 | 43.091 | 69.81 | 14.03 |
| Al/ITO/STO/$NaSnCl_3$/$MoTe_2$/Au | 0.4663 | 43.109 | 69.82 | 14.03 |
| Al/ITO/$WS_2$/$NaSnCl_3$/$MoTe_2$/Au | 0.468 | 43.085 | 69.94 | 14.1 |
| Al/ITO/IGZO/$NaSnCl_3$/$MoTe_2$/Au | 0.4669 | 43.082 | 69.89 | 14.06 |
| Al/ITO/ZnSe/$NaSnCl_3$/$MoTe_2$/Au | 0.4679 | 43.109 | 70 | 14.12 |
| Al/ITO/CSTO/$NaSnCl_3$/$MoTe_2$/Au | 0.4681 | 43.111 | 70.02 | 14.13 |

Table 4. Range of $N_c$ and $N_v$ of the absorber ($NaSnCl_3$) for higher PV performance parameters of the six structures

| Parameters | Effective density of states | Structures | | | | | |
|---|---|---|---|---|---|---|---|
| | | $SnS_2$-$MoTe_2$ | STO-$MoTe_2$ | $WS_2$-$MoTe_2$ | IGZO-$MoTe_2$ | ZnSe-$MoTe_2$ | CSTO-$MoTe_2$ |
| $V_{oc}$ | $N_c$ (cm$^{-3}$) | $10^{16}$ - $10^{17}$ | $10^{16}$ - $10^{17}$ | $10^{16}$ - $10^{17}$ | $10^{16}$ - $10^{17}$ | $10^{16}$ - $10^{17}$ | $10^{16}$ - $10^{17}$ |
| | $N_v$ (cm$^{-3}$) | $10^{16}$ - $10^{17}$ | $10^{16}$ - $10^{17}$ | $10^{16}$ - $10^{17}$ | $10^{16}$ - $10^{17}$ | $10^{16}$ - $10^{17}$ | $10^{16}$ - $10^{17}$ |
| $J_{sc}$ | $N_c$ (cm$^{-3}$) | $10^{18}$ – $10^{20}$ | $10^{18}$ – $10^{19}$ | $10^{17}$ - $10^{18}$ | $10^{18}$ – $10^{20}$ | $10^{18}$ – $10^{19}$ | $10^{19}$ – $10^{20}$ |
| | $N_v$ (cm$^{-3}$) | $10^{19}$ – $10^{20}$ | $10^{19}$ – $10^{20}$ | $10^{19}$ – $10^{20}$ | $10^{19}$ – $10^{20}$ | $10^{19}$ – $10^{20}$ | $10^{19}$ – $10^{20}$ |
| FF | $N_c$ (cm$^{-3}$) | $10^{16}$ – $10^{17}$ | $10^{16}$ – $10^{17}$ | $10^{20}$ | $10^{16}$ – $10^{17}$ | $10^{16}$ – $10^{17}$ | $10^{19}$ |
| | $N_v$ (cm$^{-3}$) | $10^{16}$ – $10^{17}$ | $10^{16}$ – $10^{17}$ | $10^{16}$ – $10^{17}$ | $10^{16}$ – $10^{17}$ | $10^{16}$ – $10^{17}$ | $10^{16}$ – $10^{17}$ |
| PCE | $N_c$ (cm$^{-3}$) | $10^{16}$ | $10^{16}$ | $10^{16}$ | $10^{16}$ | $10^{16}$ | $10^{16}$ – $10^{20}$ |
| | $N_v$ (cm$^{-3}$) | $10^{16}$ | $10^{16}$ | $10^{16}$ | $10^{16}$ | $10^{16}$ | $10^{16}$ |

Table 5. Performance parameters of the optimized structures of this work and comparison with previously reported study

| Device structure | $V_{oc}$ (V) | $J_{sc}$ (mA/cm$^2$) | FF (%) | PCE (%) | Reference |
|---|---|---|---|---|---|
| Al/ITO/$SnS_2$/$NaSnCl_3$/$MoTe_2$/Au | 1.1955 | **35.82** | **89.72** | **38.42** | This work |
| Al/ITO/STO/$NaSnCl_3$/$MoTe_2$/Au | 1.1954 | 31.45 | 89.02 | 33.46 | This work |
| Al/ITO/$WS_2$/$NaSnCl_3$/$MoTe_2$/Au | **1.2024** | 34.39 | 82.48 | 34.11 | This work |
| Al/ITO/IGZO/$NaSnCl_3$/$MoTe_2$/Au | 1.1951 | 34.62 | 89.53 | 37.04 | This work |
| Al/ITO/ZnSe/$NaSnCl_3$/$MoTe_2$/Au | 1.1954 | 34.79 | 89.6 | 37.27 | This work |
| Al/ITO/CSTO/$NaSnCl_3$/$MoTe_2$/Au | 1.1957 | 34.88 | 79.67 | 33.23 | This work |
| FTO/$TiO_2$/$NaSnCl_3$/Spiro-OMeTAD/Au | 1.17 | 34.46 | 89.33 | 36.14 | [47] |